\documentclass[prd,twocolumn,showpacs,superscriptaddress,nofootinbib]{revtex4-1}
\usepackage{amsfonts}
\usepackage{tikz}
\usepackage{pgfplots}
\usepackage{amsmath}
\usepackage{amssymb}
\usepackage{bm}
\usepackage{dcolumn}
\usepgfplotslibrary{groupplots}
\usepackage{wrapfig}
\usepackage{graphicx}
\usepackage{graphics}
\usepackage{latexsym}
\usepackage{rotating}
\usepackage[colorlinks=true]{hyperref}
\usepackage[all]{hypcap} 
\usepackage{xspace} % Sensible space treatment at end of simple macros
\usepackage{xcolor}
\usepackage{mathrsfs}
\usepackage{siunitx}
\usepackage{multirow}
\usepackage{soul}
\usepackage{lipsum}
\usepackage{colortbl}
\usepackage{ulem}
\usepgfplotslibrary{fillbetween}
\definecolor {darkgreen}{rgb}{0.2,0.7,0.2}

\newcommand\be{\begin{equation}}
\newcommand\ee{\end{equation}}
\newcommand\bw{\begin{widetext}}
\newcommand\ew{\end{widetext}}

\newcommand{\writeme}[1]{\textcolor{cyan}{\textbf{OPEN: TO BE WRITTEN\\}} }

\newcommand{\bea}{\begin{eqnarray}}
\newcommand{\eea}{\end{eqnarray}}

\newcommand{\THC}{{\tt THC\_M1}}
\newcommand{\model}[2]{$\text{#1}_{q=#2}$}

\newcommand{\cm}{{\rm cm}}

\newcommand{\g}{{\rm g}}

\usepackage{cellspace} %
\newlength\figureheight
\newlength\figurewidth

\newlength\figureheightmed
\newlength\figurewidthmed

\newlength\figureheightlarge
\newlength\figurewidthlarge

\begin{document}
\title{Impact of moment-based, energy integrated neutrino transport on 
microphysics and ejecta in binary neutron star mergers}

\author{Pedro L. Espino}
\affiliation{Institute for Gravitation \& the Cosmos, The Pennsylvania 
State University, University Park PA 16802, USA}
\affiliation{Princeton Gravity Initiative, Jadwin Hall, Princeton University, 
Princeton, NJ 08540, USA}
\affiliation{Department of Physics, University of California, Berkeley, 
CA 94720, USA}
\author{David Radice}
\thanks{Alfred P.~Sloan Fellow}
\affiliation{Institute for Gravitation \& the Cosmos, The Pennsylvania 
State University, University Park PA 16802, USA}
\affiliation{Department of Physics, The Pennsylvania State University, 
University Park PA 16802, USA}
\affiliation{Department of Astronomy \& Astrophysics, The Pennsylvania 
State University, University Park PA 16802, USA}
\author{Francesco \surname{Zappa}}
\affiliation{Theoretisch-Physikalisches Institut, Friedrich-Schiller-Universit{\"a}t Jena, 07743, Jena, Germany}
\author{Rossella \surname{Gamba}}
\affiliation{Institute for Gravitation \& the Cosmos, The Pennsylvania 
State University, University Park PA 16802, USA}
\affiliation{Department of Physics, University of California, Berkeley, 
CA 94720, USA}
\affiliation{Theoretisch-Physikalisches Institut, Friedrich-Schiller-Universit{\"a}t Jena, 07743, Jena, Germany}
\author{Sebastiano \surname{Bernuzzi}}
\affiliation{Theoretisch-Physikalisches Institut, Friedrich-Schiller-Universit{\"a}t Jena, 07743, Jena, Germany}

%%%%%%%%%%%%%%%%%%%%%%%%%%%%%%%%%%%%%%%%%%%%%%%%
\begin{abstract}
We present an extensive study of the effects of neutrino transport in 3-dimensional 
general relativistic radiation hydrodynamics (GRHD) simulations of binary neutron 
star (BNS) mergers using our moment-based, 
energy-integrated neutrino radiation transport (M1) scheme. We consider a total of 
8 BNS configurations, while varying 
equation of state models, mass ratios and grid resolutions, for a total of 16 
simulations. We find that M1 neutrino transport is crucial in modeling the local 
absorption of neutrinos and the deposition of lepton number throughout the medium. 
We provide an in-depth look at 
the effects of neutrinos on the fluid dynamics and luminosity 
during the late inspiral and post-merger phases,
%OOE 
%out-of-equilibrium dynamics and the potential impact of neutrino absorption on
%mitigating bulk viscosity in the post-merger phase, 
the properties of ejecta and outflow, and the post-merger nucleosynthesis. 
%%\dr{I would remove the numbers in the previous sentence, because they
%%give the impression that we have only these 4 results, but then the
%%abstract goes on about other things.} \ple{Fixed}
The simulations presented in this work comprise an extensive study of the combined 
effect of the equation of state and M1 neutrino transport in GRHD simulations of BNS 
mergers, and establish that the solution provided by our M1 scheme is robust 
across system properties and provides insight into the effects of neutrino trapping 
in BNS mergers.
%%\fz{I would add something about the main results. For sure I would say that this work 
%%shows that
%%properly including the neutrino component in the remnant core reduces out-of-
%%equilibrium.}
%%\ple{This has been re-written above to avoid repetition and state some of the main 
%%results without going into too much detail.}
\end{abstract}

\date{\today} \maketitle
%\tableofcontents

%%\fz{General suggestions:
%%\begin{itemize}
%%\item Add more ticks in figures \ple{Any specific figures you recommend?}
%%\item Reduce hspace and wspace in multifigures, where possible \ple{I've done this 
%%where possible, but let me know if there are specific images where you think I can do 
%%this.}
%%\end{itemize}
%%}
%===============================================================
\section{Introduction}
%===============================================================
As we approach the era of precision gravitational wave 
astronomy~\cite{LISA:2017pwj, Sathya2019, 
Maggiore:2019uih, Adhikari:2019zpy, NASALISAStudyTeam:2020lee, 
LIGOScientific:2021djp, Ballmer:2022uxx, BertiSM22, Buonanno:2022pgc}, 
the need 
for high-accuracy numerical simulations of binary neutron star (BNS) 
mergers becomes ever relevant~\cite{Duez2019, Tsokaros:2021tsu, 
Foucart:2022iwu}. 
There is presently a concerted effort to  
improve the accuracy and proper treatment of relevant physical phenomena 
in numerical relativity (NR)
codes, including: (1) the accuracy associated with 
gravitational waves (GWs)
extracted from simulations, which 
inform matched-filtered searches in current 
and future GW 
detectors~\cite{Chu:2015kft, Bernuzzi:2016pie, Most:2019kfe,
Westernacher-Schneider:2020tlr, PhysRevD.102.104014, Dudi:2021wcf, 
Doulis:2022vkx}; (2) the methods 
used for treating radiation 
and neutrino transport~\cite{Sekiguchi_2015,Radice:2018pdn,
Vsevolod:2020pak,Vincent:2019kor, Weih_2020, Foucart_2020, Radice:2021jtw}; 
(3) the accurate treatment of magnetic field 
effects~\cite{prs15,Ruiz:2017due,Ruiz_2016,Kiuchi:2017zzg,
Shibata:2021bbj,2020ApJ...900L..35C,Mosta:2020hlh,
Ciolfi_2020}; and (4) the neutron star (NS) equation of 
state (EOS)~\cite{Abdikamalov:2008df, 
Bauswein:2013jpa, Palenzuela_2015, Sekiguchi_2015, 
Lehner:2016lxy, Radice:2018pdn, Most:2018eaw, Dietrich:2018phi, 
Vincent:2019kor}. NR simulations of BNS mergers 
stand as our best tools for understanding the complex interactions of all of 
the aforementioned phenomena and effects during different stages of the 
merger process. However, substantial work remains to be done to improve the 
accuracy and microphysics within NR codes~\cite{Foucart:2022iwu}. 

An area of particular interest is the accurate treatment of neutrino 
transport in NR simulations. Different stages of a BNS merger are 
expected to produce intense neutrino emission. 
A burst of relatively low energy neutrinos (with energy 
$E_\nu \sim \mathcal{O}(\text{MeV})$) is expected during the formation of 
the central remnant~\cite{EiLiPiSc, Rosswog:2003rv}. 
Higher energy neutrinos 
%\dr{Usually people think of TeV neutrinos as
%high-energy. Ice-cube does not do well at GeV energy and traditional
%neutrino experiments (SuperK, DUNE) are focusing on MeV energies.}\ple{Changed 
%`high-energy' to `higher energy'}
$E_\nu \sim \mathcal{O}(\text{TeV})$ 
%\dr{I would say TeV, rather than GeV}\ple{Fixed} 
may also be produced by hadronic 
interactions within the relativistic jet that forms after the 
merger~\cite{Fang:2017tla, Kimura:2018vvz}.
Thermal neutrinos ($E_\nu \sim \mathcal{O}(\text{MeV})$) are 
also expected from the cooling of long-lived (with lifetimes longer 
than $t\sim\mathcal{O}(\text{s})$), post-merger 
NS remnants~\cite{Kyutoku_2018}. It may even be possible to detect 
very-high energy ($E_\nu \sim \mathcal{O}(\text{EeV})$) neutrinos on a 
timescale of days to weeks after merger, in the event that a long-lived 
magnetar remnant is produced~\cite{PhysRevD.79.103001, Fang:2017tla}. 
The peak neutrino luminosity during a BNS merger is typically 
$L_\nu \sim 10^{53}-10^{54} \SI{}{erg\per\s}$, which is a few times 
greater than that associated with core-collapse supernovae (CCSNe)
~\cite{Ruffert:1996by, Rosswog:2003rv, Sekiguchi_2015, 
Palenzuela_2015, Foucart_2016,PhysRevD.93.044019, 
PhysRevD.96.123015,PhysRevD.102.103015,Burrows:2019zce,Kullmann:2021gvo, 
Cusinato:2021zin, Radice:2021jtw}.
%% and 
%%may produce detectable neutrino fluxes in current and future Earth-based 
%%neutrino 
%%detectors~\cite{ANTARES:2017bia, Kyutoku_2018, deWasseige:2021euk, 
%%Burns:2019byj, IceCube:2021ddq}.
%%\dr{I am skeptical that we can detect MeV energy neutrinos. Our code
%%does not handle non-thermal neutrinos, which are more promising, but
%%this work cannot say much about them.}\ple{I am just listing phenomena across the 
%%neutrino spectrum that may be relevant for BNS mergers generally. Please let me 
%%know if I should remove this discussion.}
%%\dr{I would not say that they are detectable. The detection horizon is
%%at most 1 Mpc even in the most optimistic case, meaning event rate
%%1:10$^5$ years\ldots.}

During different stages of the merger, regions of the system can be in 
a state that is anywhere from the optically thin regime (wherein 
neutrinos free-stream) to the optically thick regime (wherein neutrinos 
diffuse)~\cite{Endrizzi:2019trv}. In all 
of these scenarios, due to their large energies and luminosities, 
neutrinos are expected to play an important role. The impact of neutrinos is 
especially relevant in the environment \emph{following} a BNS merger.
%%\fz{In the following, I think we should be more specific: I think that
%%this refers to the post-merger remnant core and maybe the highest density
%%regions of the disk (depending on the neutrino type and relative neutrinosphere)}
%%\ple{I've now pointed out that neutrinos are especially important in the post-merger environment.}
For example, in the diffusive limit, neutrino interactions are 
expected to change the matter composition. A changing 
matter composition is relevant for determining the conditions 
relevant for
$r$-process nucleosynthesis~\cite{Lattimer:1974slx, 
Just:2014fka, Thielemann:2017acv, Perego:2021dpw, Radice:2021jtw} 
and may also emergently lead to dissipation in out-of-equilibrium fluid 
dynamics~\cite{Alford:2017rxf, Hammond:2021vtv, Most:2022yhe, Celora:2022nbp, 
Chabanov:2023blf, Espino:2023dei}. 
%%In the free-streaming limit neutrino emission provides the dominant 
%%cooling mechanism for the \fzedit{system}{remnant} 
%%\ple{I think the word `system' better captures all scenarios (such as BH+disk and RMNS)}. 
%%\dr{I am not sure of what it means to say that free-streaming neutrinos
%%provide cooling: the free-streaming neutrinos are, by definition, not
%%interacting with matter anymore.}
Neutrinos that are 
produced in the hot 
and dense regions may reach the conditions for decoupling and may 
eventually be emitted from the system, thereby carrying off 
energy~\cite{Rosswog:1998hy}. As the neutrinos remove energy from the 
system, they may lead to additional matter outflow in the form of 
neutrino-driven winds~\cite{Rosswog:1998hy,Rosswog:2002rt, 
Dessart:2008zd, Perego:2014fma, Radice:2018xqa, Vsevolod:2020pak}. 
Even in the free-streaming limit neutrino irradiation may significantly 
change the composition of the low-density regions of the remnant and 
ejecta via charged-current interactions~\cite{Dessart:2008zd, 
Perego:2014fma, Just:2014fka, Foucart_2016, Sekiguchi:2016bjd, 
Sekiguchi:2016bjd}.

Capturing the aforementioned effects in BNS mergers simulations 
requires a sufficiently accurate treatment of neutrino transport in NR 
codes, which in principle requires a solution to the Boltzmann equations 
of radiation transport. Full solutions to the Boltzmann equation require 
the evolution of a 7-dimensional distribution function for each 
species of neutrino considered, which presents a computationally 
intensive problem. Many alternative approaches for capturing the 
effects of neutrino transport have been considered in the context of BNS 
merger simulations, including both direct and approximate methods. For 
example, some modern methods that \emph{directly} tackle the solution of 
the Boltzmann equation include the expansion of momentum-space 
distributions into spherical harmonics~\cite{Radice:2012tg}, 
lattice-Boltzmann based methods~\cite{Weih_2020}, Monte Carlo (MC)-based 
methods~\cite{Abdikamalov:2012zi, Richers:2015lma, Ryan:2015mda, 
Foucart:2017mbt, Foucart:2018gis, Foucart_2020, Foucart:2021mcb}, and 
discrete-ordinates-based methods~\cite{Nagakura:2022qko}. Compared to 
approximate methods, direct methods are generally more accurate and 
suffer less from model-dependence; on the other hand, they are often 
complex to implement into numerical codes and computationally 
intensive. For instance, MC-based 
methods may become prohibitively expensive in the optically thick regime, 
where the neutrino mean-free path (which must be resolved to properly 
capture neutrino-matter thermal equilibrium) becomes small. Recent 
developments in the use of MC-based schemes address the need for 
prohibitive amount of MC particles by modifying the relevant 
interaction rates in regions of high optical depth, such that the 
neutrino energy distribution is unaffected close to regions where the 
neutrinos decouple from the matter~\cite{Foucart_2020, Foucart:2021mcb}. 
Such an approach may 
%%\fz{From now on the expression "reliably capture" and the word
%%"capture" are used
%%too many times.. three already in this very sentence} \ple{Fixed}
reliably capture the effects of neutrinos in the 
optically thin and optically thick limits but cannot do so for the 
emergent out-of-equilibrium effects captured by a fluid treatment of the 
radiation~\cite{Alford:2017rxf, Hammond:2021vtv, Most:2022yhe, Celora:2022nbp, 
Chabanov:2023blf, Espino:2023dei}.

Approximate methods to the solution of the Boltzmann 
equation model the relevant effects of neutrino transport 
in BNS mergers while reducing computational expense. For example, 
neutrino leakage schemes~\cite{Sekiguchi:2010fh, Sekiguchi:2011zd, 
stellarcollapse, Wanajo:2014wha, PhysRevD.93.044019, Radice:2016dwd, 
Perego_2017, Ardevol-Pulpillo:2018btx} are computationally 
inexpensive and may reliably capture the neutrino 
cooling effects. 
Traditional leakage 
schemes, however, cannot account for neutrino transport throughout the 
system and as such cannot provide accurate insight into higher-order 
effects, such as the deposition of heat and electron fraction throughout 
the system~\cite{PhysRevD.93.044019, Wanajo:2014wha}. 
%%\fz{Maybe add an explicity sentence about M0 and then talk about 
%%Boltzmann 7D -> 3+1 -> M1?} \ple{I think we have enough information to set the stage. 
%%Please let me know if you feel strongly about having descriptions of additional 
%%neutrino transport schemes.}
An approximate 
method at a higher level of sophistication beyond leakage is the use of 
moment-based schemes~\cite{Shibata:2011kx, Sekiguchi_2015, Foucart_2015, 
Radice:2016dwd}, in which the 7-dimensional Boltzmann 
equation 
are reduced to
%%\dr{I would say ``reduced'' since this is an approximation} 
a system of 3+1 equations similar to the equations of general 
relativistic hydrodynamics (GRHD). 
Generally, 
moment-based schemes employ an expansion in the moments of the neutrino 
distribution 
function truncated at a given order~\cite{Shibata:2011kx}, which are then 
evolved together with the fluid.
%%moment-based schemes 
%%consider the evolution of moments of the neutrino distribution function, 
%%such as the \fzedit{the}{} energy and momentum density. Moment\fz{}{-}based 
%%schemes employ an expansion in the moments of the neutrino distribution 
%%function which is truncated at a given order~\cite{Shibata:2011kx}. 
As 
such, a key feature of moment-based schemes is the need for an analytical 
closure for the transport equations, achieved by providing a form for 
moments at an order above which the expansion is truncated. The most 
recent and accurate implementations of moment-based schemes focus on the 
use of the energy-integrated (M1) scheme~\cite{Radice:2021jtw}, which 
also require an analytic estimate of the neutrino energy spectrum. Recent 
developments in the use of M1 schemes in the context of BNS merger 
simulations are able to 
capture the diffusion limit of radiation 
transport without the need for the ill-posed relativistic heat transfer 
equation~\cite{PhysRevD.31.725, Andersson:2011fd}, and retain all of the 
matter-coupling terms that appear in the evolution 
equations~\cite{Radice:2021jtw}. These advancements are crucial for 
reliably capturing the trapping of neutrinos in relativistic 
media~\cite{Radice:2021jtw}. 
%%\fz{Following sentence seems to long, consider to split. for instance between 
%%"closure used" and
%%"form assumed" you could put a period.}
%%\ple{Broke sentence up in a different way 
%%than suggested}
Despite their accuracy in the diffusion limit and relatively low 
computational expense (compared to the approaches which solve the full Boltzmann 
equations), 
moment-based schemes generally 
suffer from model-dependence in the particular closure 
used~\cite{PhysRevD.102.083017} and form assumed in the neutrino energy 
spectrum~\cite{Foucart_2016}. Moment-based schemes are also 
known to produce unphysical shocks in 
regions where radiation beams cross~\cite{Foucart:2018gis} 
%(an effect that may be addressed by the albeit more expensive lattice-Boltzmann 
%approaches~\cite{Weih_2020}) 
%\dr{I think we are giving too much credit to
%Weih et al., lattice Boltzmann is at most as good as the moment method.}
%\ple{comment removed},
and are not expected to converge to the 
solution of the Boltzmann equations~\cite{Foucart:2021mcb}. 
%%\fz{The following sentence looks a bit redundant to me.}
%%Nevertheless,
%%M1 schemes may be suitable for accurately capturing all of the relevant 
%%neutrino effects in BNS merger simualtions, without the need for the 
%%expensive treatments that directly solve the Boltzmann 
%%equation~\cite{Radice:2021jtw, Foucart:2021mcb, Weih_2020}.

Full, 3D general relativistic hydrodynamics 
%%\fz{I would remove 3D from this acronym}
simulations with
some of the aforementioned high-order methods (specifically, MC
and M1 neutrino treatments) show general agreement. Specifically, differences of 
approximately 10\% in the properties of the most sensitive (funnel) region of the 
outflow and differences of up to $20\%$ in the neutrino luminosities and energies 
arise between the two methods~\cite{Foucart_2020}. 
%%\dr{I think we
%%need to be more specific. Foucart compared the $Y_e$ in the most
%%sensitive part of the outflow (in the funnel region). For other things
%%the error due to M1 are probably smaller.}\ple{Fixed}
However, the error introduced by the M1 approximation in other quantities may be 
smaller. Additional work remains to be done which systematically compares the
results obtained using MC or full-Boltzmann based methods
with those obtained using M1 neutrino transport and other lower-order
approximate methods. Given the wide range of neutrino transport schemes,
different treatments may be better suited for different research
questions. For instance, MC or full-Boltzmann based methods
may be best suited for high-accuracy simulations to understand the
solution to which other methods ought to converge and to reliably
understand the systematic errors that may arise in the use of
approximate methods. However, their relatively high computational demand
does not make these methods the best option for parametric studies that
are designed to cover a large portion of the parameter space.
%%wide variety of simulations.
%%\fz{better say "a large portion of the parameter space"?}. 
On the other hand,
in the case of moment-based methods, we can make use of the relatively
high computational efficiency and suitable accuracy to efficiently
explore the parameter space of BNS merger simulations while accurately
capturing neutrino effects. 

In this work we employ the \THC{} code -- an
extension of the {\tt THC} code which employs an updated M1 neutrino
transport as detailed in~\cite{Radice:2021jtw} -- to run 3D 
GRHD of a wide
variety of BNS mergers. We consider binary systems across several 
equation of state models and two mass ratios, and report on relevant
observables including the gravitational and neutrino radiation,
%OOE
%out-of-equilibrium dynamics, 
ejecta properties, and nucleosynthesis,
among others. Our work extends on the case studies presented
in~\cite{Radice:2021jtw} and~\cite{Zappa:2022rpd}, and significantly
expands the catalog of results for 3D GRHD simulations of BNS mergers
with M1 neutrino transport. Our M1 neutrino treatment allows us to
improve on previous parameter studies with the {\tt THC} code, which
employed a lower-order (M0) neutrino transport scheme and neutrino
leakage. Crucially, our 
%%\fz{"energy-integrated" is a key feature of the
%%model, I would not slip it in here without explanation. Consider to add a sentence
%%about it or move this discussion when M1 is reviewed}\ple{Added a reference to the 
%%section where we discuss the scheme.} 
energy-integrated scheme, which we discuss in detail in 
Sec.~\ref{subsec:M1_equations}, allows us to accurately
determine the conditions for neutrino decoupling, as decoupling surfaces
are highly sensitive to the neutrino energy, and as such
energy-integrated schemes are expected to be more accurate in this
regard. Additionally, our large set of simulations allows us to
study the combined effects of the 
EOS and neutrino transport 
in optically thick regions.

The remainder of the work is organized as follows. In 
Sec.~\ref{sec:numerical_methods} we outline the main numerical methods 
used in this work, including a brief description of the \THC{} code and the 
diagnostics used to analyze our simulations. In 
Sec.~\ref{sec:simulations} we detail our grid setup,  
and full suite of simulations considered. 
%%\fzedit{We also discuss the 
%%grid hierarchy used for our simulations.}{} 
In Sec.~\ref{sec:results} we 
discuss the key results of our simulations, with particular focus on the 
merger dynamics, gravitational waves, %out-of-equilibrium fluid dynamics, %OOE
merger ejecta, neutrino luminosity, and nucleosynthetic yields. Finally, in 
Sec.~\ref{sec:conclusion} we summarize the main findings of our 
simulations and list the key effects of using our improved M1 neutrino 
treatment. Additionally, we consider the convergence properties of relevant 
quantities in our 
simulations in App.~\ref{app:convergence}. Throughout the work, we assume 
geometrized units, where $G=c=1$, and allow Greek (Roman) tensor indices 
%$(\alpha, \beta, \ddots)$ 
to run over four (three) dimensions, unless otherwise noted.

%In this work we consider 3D, general-relativistic hydrodynamics simulations 
%of BNS 
%mergers with the moment-based, energy-integrated (M1) neutrino transport 
%scheme detailed 
%in ...

%===============================================================
\section{Numerical methods}\label{sec:numerical_methods}
%===============================================================
%---------------------------------------------------------------
\subsection{Evolution code}
%---------------------------------------------------------------
%%\fz{I think that all this until before "We solve.." can be removed.
%%If you decide to keep it, you have to say that the spacetime is 
%%foliated into spacelike hypersurfaces, that are "connected" by lapse
%%and shift and that these elements represent the coordinate choice/
%%gauge freedom.}
%%Our simulations require the numerical solution of the Einstein field 
%%equations
%%\begin{equation}\label{eq:Einstein}
%%G_{\mu\nu} = 8\pi T_{\mu\nu},
%%\end{equation}
%%(where $G^{\mu\nu}$ and $T^{\mu\nu}$ are the Einstein and stress-energy 
%%tensors, respectively), 
%%coupled to the equations of ideal relativistic hydrodynamics. We work 
%%with the \fzedit{a}{} spacetime metric \fzedit{of the form}{}
%%\begin{equation}
%%ds^2 = -\alpha^2 dt^2 + \gamma_{ij}(dx^i + \beta^i dt)(dx^j + 
%%\beta^j dt).
%%\end{equation}
%%where $\alpha$ is the lapse, $\beta^i$ is the shift, 
%%$\gamma_{ij} \equiv g_{ij} + n_i n_j$ is the three-metric,
%%%\dr{induced is not clear here, 
%%%because you did not introduce the foliation} \ple{fixed}, 
%%and $n^i=(1/\alpha, \beta^i/\alpha)$ is the 
%%future-pointing unit vector orthogonal to each 
%%constant-$t$ spatial hypersurface. 
%where $\gamma_{ij}$ is the three-metric, $\gamma$ is the determinant of 
%the three-metric, $\alpha$ is the lapse, $\beta^i$ is the shift, $K_{ij}
%$ is the extrinsic curvature
We solve the Einstein equations with the {\tt CTGamma} code,
%%\dr{I believe that the name of the code is CTGamma}
which implements the Z4c 
formulation~\cite{Bernuzzi:2009ex,Hilditch:2012fp}
of the Einstein equations. 
Our gauge conditions consist of the ``1+log" slicing condition for the 
lapse~\cite{Bona:1994dr} and an ``integrated Gamma-driver" condition for 
the shift~\cite{Alcubierre:2002kk}, with the shift coefficient set to 
$\nu=0.75$ and damping coefficient set to $\eta=2.0$. 
Time-integration is carried out using a third-order accurate Runge-Kutta 
(RK3) scheme, using the method of lines with the {\tt MoL} thorn, 
with a Courant factor of 0.15.
We solve the equations of relativistic radiation-hydrodynamics 
with the {\tt THC\_M1} code, which is an extension of the {\tt THC} 
code that includes the M1 moment-based neutrino treatment described 
in the following section (Sec.~\ref{subsec:M1_equations}). 
Additionally, we model subgrid-scale 
viscous angular momentum transport using the general-relativistic large-eddy 
simulation (GRLES) formalism~\cite{Radice:2020ids}. We leave the settings of the 
LES model fixed for all simulations.

%---------------------------------------------------------------
\subsection{Moment-based, energy-integrated neutrino transport}\label{subsec:M1_equations}
In this work we treat neutrino radiation transport within the M1 scheme, 
which describes the neutrino fields in terms of their energy-integrated 
stress energy tensors. We consider 3 distinct neutrinos species including
the electron neutrino $\nu_{\rm e}$, electron anti-neutrino 
$\bar{\nu}_{\rm e}$, and heavy-lepton species neutrinos, grouped 
into $\nu_{\rm x}$. For each neutrino species, the stress-energy tensor 
takes the form 
%%\fz{I would remove NR for consistency with Radice+2022}\ple{Doing it this way 
%%allows me to distinguish between the radiation and hydro stress-energy tensors, and 
%%it's still clear what is meant by each term.}
\begin{equation}\label{eq:stressenergy_euler}
T_{\rm NR}^{\alpha\beta} = En^\alpha n^\beta + F^\alpha n^\beta + 
n^\alpha F^\beta  + P^{\alpha\beta}, 
\end{equation}
%where $n^\alpha$ is the unit-normal oriented orthogal to constant-$t$ 
%hypersurfaces, 
where $E$ is the radiation energy density, $F^\alpha$ is the radiation 
flux, and $P^{\alpha \beta}$ is the radiation pressure tensor in the 
Eulerian frame. We note that $F^\alpha n_\alpha = 0$ and 
$P^{\alpha\beta}n_\alpha = 0$. In the fluid rest frame, we can write the 
quantity in Eq.~\eqref{eq:stressenergy_euler} as
\begin{equation}\label{eq:stressenergy_fluid}
T_{\rm NR}^{\alpha\beta} = Ju^\alpha u^\beta + H^\alpha u^\beta + u^\alpha 
H^\beta + K^{\alpha\beta},
\end{equation}
where $u^\alpha$ is the fluid four-velocity, $J$ is the radiation energy 
density, $H^\alpha$ is the radiation flux, and $K^{\alpha \beta}$ is the 
radiation pressure tensor in the fluid rest frame. We note that 
conservation of energy and angular momentum requires that 
\begin{equation}\label{eq:conservation_stressenergy}
\nabla_\beta T^{\alpha\beta}_{\rm NR} = -\nabla_\beta T^{\alpha\beta}_{\rm HD},
\end{equation}
where $T^{\alpha\beta}_{\rm HD}$ is the matter stress-energy tensor. In 3+1 form, 
Eq.~\eqref{eq:conservation_stressenergy} takes the form
%%\fz{I think in the second term a factor $\sqrt{\gamma}$ is missing, cmp with Radice+2022}\ple{Good catch!}
\begin{eqnarray}\label{eq:3p1_conservation}
\partial_t(\sqrt{\gamma}E) + \partial_i\left[\sqrt{\gamma}(\alpha F^i - \beta^i E)\right] = \nonumber \\
\alpha\sqrt{\gamma} \left[ P^{ik}K_{ik} - F^i\partial_i \log\alpha - \mathcal{S}^\mu n_\mu \right], \nonumber \\
\partial_t(\sqrt{\gamma} F_i) + \partial_k\left[\sqrt{\gamma}\left(\alpha P^k_i - \beta^k F_i \right)\right] = \nonumber \\
\sqrt{\gamma} \left[-E\partial_i \alpha + F_k\partial_i \beta^k + \frac{\alpha}{2} P^{jk} \partial_i \gamma_{jk} + \alpha \mathcal{S}^\mu \gamma_{i\mu} \right],
\end{eqnarray}
where $K_{ik}$ is the extrinsic curvature and the term $\mathcal{S}^\mu$, which takes the form \cite{Shibata:2011kx}
\begin{equation}\label{eq:matter_radiation_interaction}
\mathcal{S}^\mu = (\eta - \kappa_{\rm a} J ) u^\mu - (\kappa_{\rm a} + \kappa_{\rm s}) H^\mu,
\end{equation}
contains the interaction terms between the neutrinos and the fluid; in 
Eq.~\eqref{eq:matter_radiation_interaction} $\eta$ is the neutrino 
emissivity, and $\kappa_{\rm a}$ and $\kappa_{\rm s}$ are the absorption 
and scattering coefficients, respectively. We assume that scattering is 
isotropic and elastic.

Eqs.~\eqref{eq:3p1_conservation} require a closure to be solved. 
Generally, M1 schemes call for an approximate analytic closure of the 
form $P^{ik} = P^{ik}(E, F^i)$. In {\tt THC\_M1}, we employ the Minerbo 
closure 
%%, which has exact forms in the optically thick and thin regimes (in the 
%%presence of spherical or slab symmetry),
%%\dr{I would not say that it is exact in the thin regime. It is not
%%really true (although a lot of people say that).} \fz{maybe here you can add 
%%"in presence of spherical or slab symmetry"..?} \ple{Is the statement now correct?} 
%%\dr{Technically we assume axial symmetry of the neutrino distribution
%%function, not spherical or slab symmetry. That said, I do not think we
%%need to state this at all, as it is inconsequential for the rest of the
%%paper.}
which takes the form
\begin{equation}
P_{\alpha\beta} = \dfrac{3\chi - 1}{2}P_{\alpha\beta}^{\rm thin} + \dfrac{3(1 - \chi)}{2}P_{\alpha\beta}^{\rm thick},
\end{equation}
where $P_{\alpha\beta}^{\rm thin}$ and $P_{\alpha\beta}^{\rm thick}$ are the closure forms in the optically thing and thick regimes, respectively, $\chi$ is the Eddington factor, given by 
\begin{equation}
\chi = \frac{1}{3} + \xi^2 \left(\dfrac{6\xi^2 -2\xi + 6}{15}\right),
\end{equation}
and
\begin{equation}
\xi^2 = \dfrac{H^\alpha H_\alpha}{J^2}.
\end{equation}
In optically thin regions $\xi \approx 1$ and $\chi \approx 1$, so 
$P_{\alpha\beta} \approx P_{\alpha\beta}^{\rm thin}$. On the other hand, in the optically thick regime $H^\alpha \approx 0$ and $\chi \approx \frac{1}{3}$, so $P_{\alpha\beta} \approx P_{\alpha\beta}^{\rm thick}$.
We refer the reader to ~\cite{Radice:2021jtw} for further details on the specific forms of $P_{\alpha\beta}^{\rm thin}$ and $P_{\alpha\beta}^{\rm thick}$, and for details on the numerical implementation of Eqs.~\eqref{eq:3p1_conservation} within {\tt THC\_M1}.
\label{subsec:M1_equations}
%---------------------------------------------------------------

%---------------------------------------------------------------
\subsection{Diagnostics}\label{subsec:diagnostics}
%---------------------------------------------------------------
We use several diagnostics to assess the state of our simulations and 
extract meaningful physical results. In the following we provide details 
on the main diagnostics used in our simulations. Where relevant, we 
highlight the specific codes and numerical methods used to report our 
findings.

To monitor collapse, we 
consider the evolution of the minimum of the lapse 
function $\alpha_{\rm min}$. We treat the threshold 
$\alpha_{\rm min} \leq 0.2$ as indicative of gravitational collapse to a 
black hole (BH). We also periodically search our numerical grid for the 
existence of an apparent horizon (AH) using the {\tt AHFinderDirect} 
code~\cite{Thornburg:2003sf}, which searches for the outermost marginally 
trapped surface on each spacelike hypersurface. 
%%\bs{. [I'd remove the following, we do not need the eq. Better to show only the 
%%relevant eqs]}\ple{Done}
%%, by solving the equation
%%\begin{equation}
%%\Theta \equiv \nabla_i n^i + K_{ij}n^in^j + K = 0,
%%\end{equation}
%%where $K=K_i^i$.

We extract GWs at the surface of fixed concentric 
spheres (centered on the origin) of several radii and report the values 
in the wave zone, which corresponds to an extraction radius of 
%%\fz{usually waves in THC are extracted at 400 M, which corresponds to 592 km.
%%The number you give here would correspond to 300 M, is this correct?}\ple{Fixed}
$r_{\rm ex}=\SI{592}{km}$. 
In particular, we use the {\tt WeylScal4} 
code~\cite{Loffler2012ETK}, 
which works within the Newman-Penrose 
formalism,~\cite{Newman:1961qr, Penrose:1962ij} and compute the 
coefficients of the $s=-2$ spin-weighted spherical harmonic 
decompositions of the Newman-Penrose scalar $\Psi_4$ using the {\tt Multipole} 
code~\cite{Loffler2012ETK}.
These coefficients are labeled as $\Psi_4^{l,m}$, where $l$ and $m$ are 
the degree and order of the spherical harmonics, respectively. 
Where relevant, we compute 
the GW strain $h$ as
\begin{equation}
\Psi_4 = \ddot h_+ - i \ddot h_\times,
\end{equation}
using the fixed-frequency integration (FFI) method~\cite{Reisswig_2011GWs}. 
Finally, we approximate the 
merger time $t_{\rm mer}$ as the time when the GW strain amplitude
reaches its peak value.

We consider several global quantities to monitor the merger fluid 
dynamics, radiation dynamics and ejecta properties.
%We compute the disk mass as
%\begin{equation}
%M_{\rm disk} = \int \sqrt{\gamma} W \rho d^3x,
%\end{equation}
%where $\gamma$ is the determinant of spatial metric and $W$ is the 
%Lorentz factor. Furthermore, we impose a cutoff on the rest mass density 
%and lapse such that we consider as part of the disk only matter with 
%$\rho_{\rm disk} < 10^{13}\SI{}{\g\per\cm\cubed}$ (as we conventionally 
%identify the remnant as regions with 
%$\rho_{\rm disk} < 10^{13}\SI{}{\g\per\cm\cubed}$) and 
%$\alpha_{\rm disk} \geq 0.3$ (to ensure that all disk material is 
%outside of the AH), respectively. 
For a qualitative understanding of the 
neutrino radiation dynamics, we consider the evolution of the neutrino 
luminosity $L_{\nu}$ and its dependence on other relevant quantities. For 
an understanding of ejecta properties, we consider the flux of matter on 
a coordinate sphere of radius $r\approx \SI{440}{km}$, and
classify fluid elements as unbound based on the Bernoulli criterion, such that fluid 
elements with 
%%\dr{I think that we actually use $h u_t < - h_{\min}$. It should not
%%matter because our EOSs are normalized to have $h_{\min} = 1$. However,
%%we should comment on this, since Foucart 2022, cited below, makes a
%%point that one introduces an error with Bernoulli, but we actually do
%%the right thing.} \ple{Fixed the discussion.}
\begin{equation}\label{eq:bernoulli}
h u_t < -h_{\rm min}
\end{equation}
are labeled as ejecta, where $h$ is the specific enthalpy, $u_t$ is the temporal 
component of the 4-velocity, and $h_{\rm min}$ is the minimum value of the specific 
enthalpy available in the tabulated EOS models we employ. We note that use of the 
Bernoulli criterion is expected to overestimate the amount of 
ejecta by assuming that all internal energy is 
converted to kinetic energy in the fluid element~\cite{Foucart:2022kon}.
%%Although 
%%the criterion shown in Eq.~\ref{eq:bernoulli} would avoid this issue, we note that 
%%the EOS models we employ in our simulations are normalized to have 
%%$h_{\rm min} = 1$.
%%\dr{There is some confusion here. The Bernoulli criterion always
%%overestimates the ejecta mass, because it assumes conversion of internal
%%to kinetic energy. Using $h_{\rm min}$ accounts for the fact that with
%%realistic EOS the minimum of the specific enthalpy is not $1$ (but it is
%%with our definitions of the EOSs).}\ple{I've removed part of the statement}
Nevertheless, the use of Eq.~\eqref{eq:bernoulli} provides a 
reasonable estimate for ejecta properties.
We consider histograms of the ejecta mass for several 
relevant fluid variables, including the electron fraction $Y_{\rm e}$, specific 
entropy $s$, temperature $T$, and asymptotic speed 
$v_\infty = \sqrt{2[h(E_\infty + 1) -1]}$. We calculate 
nucleosynthetic yields following the procedure highlighted in~\cite{Radice:2018pdn} 
using the {\tt SkyNet} code~\cite{Lippuner_2017}. We also compute synthetic 
kilonova (KN) light 
curves following the procedure highlighted in~\cite{Wu:2021ibi} using the {\tt SNEC} 
code~\cite{Morozova:2015bla}.
%%Specifically, we calculate the absolute bolometric (AB) 
%%magnitude using
%%\dr{Do we need this equation?}
%%\begin{equation}
%%m_{\rm AB} = -2.5 \log\left( \dfrac{\int f_\nu(h\nu)^{-1} e(\nu) d\nu}{\int 3631{\rm 
%%Jy}(h\nu)^{-1} e(\nu)d\nu} \right),
%%\end{equation}
%%where $\nu$ are $f_\nu$ the light frequency and flux density assuming a source of 
%%$\SI{40}{Mpc}$ and $e(\nu)$ are filter functions for the different bands of the 
%%Gemini instrument~\cite{Wu:2021ibi}.

To monitor the dynamics of the fluid at different stages of the merger, 
we consider the evolution of several fluid variables on the equatorial 
and meridional planes, with particular focus given to the rest mass 
density 
$\rho$, temperature $T$, electron fraction $Y_{\rm e}$, 
and electron neutrino neutrino radiation energy in the lab frame
$E_{\nu_{\rm e}}$. 

\section{Simulations}\label{sec:simulations}
%===============================================================
\begin{table}[htb]\label{tab:configurations}
\centering
\caption{Summary of the BNS configurations considered in this work. We 
list: the EOS; the central value of the specific enthalpy $h_{\rm c}$, 
which is relevant for the construction of initial data with the {\tt 
Lorene} code; the mass ratio $q=M_1/M_2$ where $M_1$ ($M_2$) is the mass of the 
more (less) massive star in the configuration, binary
mass $M$,
%%\bs{[usually we keep $M$ for the binary mass, and use $M_{\rm ADM}$ for the 
%%initial ADM mass. I think here you mean binary mass, not initial ADM mass.]}
%%\ple{Fixed}
and total system baryonic mass $M_{\rm b}$; the maximum mass of a 
non-rotating star for the given EOS $M_{\rm max}^{\rm TOV}$; 
the grid resolutions considered (where LR and SR 
stand for low and standard resolution, respectively); and
the label used to refer to each model.
Our full set of simulations consists of 16 simulations.
%%\fz{Here I would remove the column $h_c$ and $M$ and just give the common value.}
%%\ple{I think it helps if other \\
%%people want to consider the same systems we are studying}
%%\dr{Why is the total mass not 2.7?}\ple{Fixed. I think the total mass is actually something like 1.345 + 1.345, and I was not rounding}
}
\begin{ruledtabular}
\begin{tabular}{l | c | c c c | c | c | c } %% \hline  \hline
EOS  & $h_{\rm c}$ & $q$ & $M $  & $M_{\rm b} $ &
$M_{\rm max}^{\rm TOV}$ & Res. & Name\\ 
 & & & $(M_\odot)$ & $(M_\odot)$ & $(M_\odot)$ & \\\hline
BLh  & 0.125 & 1.0 & 2.7 & 2.95 & 2.10 & LR/SR & \model{BLh}{1} \\ \hline
BLh  & 0.125 & 1.2 & 2.7 & 2.95 & 2.10 & LR/SR & \model{BLh}{1.2} \\ \hline
DD2  & 0.125 & 1.0 & 2.7 & 2.94 & 2.48 & LR/SR & \model{DD2}{1} \\ \hline
DD2  & 0.125 & 1.2 & 2.7 & 2.94 & 2.48 & LR/SR & \model{DD2}{1.2} \\ \hline
SFHo & 0.125 & 1.0 & 2.7 & 2.96 & 2.06 & LR/SR & \model{SFHo}{1} \\ \hline
SFHo & 0.125 & 1.2 & 2.7 & 2.96 & 2.06 & LR/SR & \model{SFHo}{1.2} \\ \hline
SLy  & 0.125 & 1.0 & 2.7 & 2.97 & 2.06 & LR/SR & \model{SLy}{1} \\ \hline
SLy  & 0.125 & 1.2 & 2.7 & 2.97 & 2.06 & LR/SR & \model{SLy}{1.2} \\ %% \hline
%SRO*  & 1.0   & 2.57472 & 2.84262 \\ \hline
\end{tabular}
\end{ruledtabular}
\end{table}
In the following we outline the main simulations considered in this work. 
We detail the properties of the initial data and discuss the grid 
setup in our simulations.
%---------------------------------------------------------------
\subsection{Equations of state and initial configurations}
\label{subsec:EOS_ID}
%---------------------------------------------------------------
\begin{figure}[htb]
\centering
\includegraphics{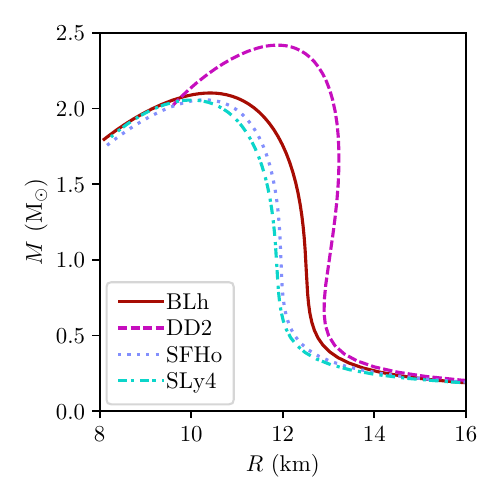}
\caption{TOV sequences for the EOS models considered in this work. 
%%\fz{Consider to add pulsar lines, see \url{http://www.computational-relativity.org/eos/}}\ple{Is it really necessary?}
}
\label{fig:TOVs}
\end{figure}

\begin{figure*}[htb]
\centering
\includegraphics[width=\textwidth]{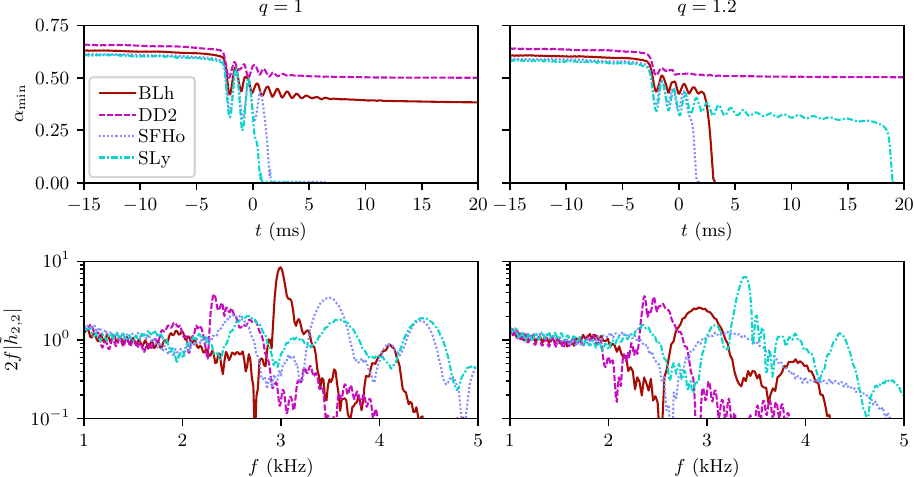}
\caption{
%%\bs{[here we need a short, main description of the figure]} \ple{Done.}
Representative global quantities for the standard resolution (SR) simulations in our 
work. {\it Top panel}: Minimum of 
the lapse function $\alpha_{\rm min}$ in the case of SR simulations with 
equal (left panel) and unequal (right panel) mass ratios. 
{\it Bottom panel}: 
GW spectra for the same simulations depicted in the top panel.
%%\dr{I would suggest to plot lapse as a function of $t - t_{\rm mrg}$.}
%%\dr{The green and yellow bands seem to be a bit arbitrary. I would have
%%  the yellow band span the range of 1-4 kHz.}
%%\bs{[I'd remove the bands, the inspiral-merger is a continuous process, there is no separation frequency.
%%    it is also easy to identify by eyes the postmerger part. At most I would put a vertical line at the 22-amplitude's peak, the ``moment of merger'']}
%%\bs{[bottom: the spectra at the right of the green band are unphysical, that is just the FT window related to the finite initial separation, They should be removed. The spectra should start at the minimum phys frequency of the initial data.]}\ple{Fixed}
}
%%\caption{Global diagnostics as functions of time for a set of 
%%representative simulations in our work. In the top panel we show the 
%%maximum rest mass density $\rho_{\rm max}$, scaled by its value at the 
%%start of each simulation in the case of equal (left panel) and unequal
%%(right panel) mass ratios. In the bottom panel we show the minimum of 
%%the lapse function $\alpha_{\rm min}$ in the case of equal (left panel) 
%%and unequal (right panel) mass ratios. We show results corresponding to 
%%the BLh, DD2, SFHo, SLy, and SLy4 EOS models using solid maroon, dashed 
%%magenta, dotted, purple, dash-dotted blue, and dash-dot-dotted green 
%%lines, respectively. All results depicted are for standard resolution 
%%(SR) simulations.
%%}
\label{fig:rhomax_alphmin}
\end{figure*}

The EOS models we consider consist of the BLh~\cite{Bombaci:2018ksa}, 
DD2~\cite{Hempel2010}, 
SFHo~\cite{Steiner_2010}, and
SLy4~\cite{Chabanat:1997un, PhysRevC.96.065802} models. 
In Fig.~\ref{fig:TOVs}, we show the sequences of static, non-rotating 
equilibrium stars (i.e., the Tolman-Oppenheimer-Volkoff (TOV) sequence)  
corresponding to each EOS considered in this work. We show the TOV 
sequences for the BLh, DD2, SFHo, and SLy EOSs using the solid maroon, 
dashed magenta, dotted blue, and dash-dotted cyan lines, respectively.
All EOS models produce maximum mass TOV 
stars with $M_{\rm max}^{\rm TOV} \geq \SI{2}{M_\odot}$, and as such are 
consistent with massive pulsar 
observations~\cite{Cromartie:2019kug, Antoniadis2013,
antoniadis2016millisecond}. Several recent, independent methods
constrain the radius of a $M=\SI{1.4}{M_\odot}$ star to 
$\SI{10}{km} \lesssim R_{1.4} \lesssim \SI{13}{km}$, including 
observations of pulsars in globular clusters~\cite{2018MNRAS.476..421S}, 
observations of x-ray pulsars~\cite{Ozel_2016mr, Bogdanov_2019I}, the 
GW170817 event~\cite{Radice:2018ozg, Christian:2018jyd, Malik:2018zcf, 
Abbott:2018exr, Landry:2018prl, Carson:2018xri, Essick:2019ldf, Raithel:2019uzi}, and 
the recent NICER results~\cite{Raaijmakers:2021uju,Miller:2021qha,
Riley:2021pdl}. All EOS models that we consider obey 
these constraints on $R_{1.4}$.
%%The DD2 
%%EOS predicts $R_{1.4} \approx \SI{13.2}{km}$, and as such is consistent 
%%with most of the aforementioned independent constraints 
%%on $R_{1.4}$; the DD2 EOS is the stiffest EOS in our set and we include it
%%to better understand the effects of stiff EOS models and to widen the 
%%diversity of the EOSs we consider.
%%\dr{I think DD2 satisfy GW and NICER constraints very well. The only
%%constraints that are violated as those for X-ray estimates of radii, but
%%those are possibly problematic.}\ple{Removed the statement about DD2}

We note that all EOS models considered in this work contain strictly 
\emph{hadronic} degrees of freedom. EOS models of this type have strong 
universality properties, and typically exhibit a $\sim 20\%$ increase
in the maximum mass when allowing for maximal uniform 
rotation~\cite{Breu2016}, a limit referred to as the `supramassive' 
mass $M_{\rm supra}$. If the total system mass falls below $M_{\rm supra}$, we
expect that the post-merger remnant will not collapse to a BH.
%%In the absence of prompt BH formation, $M_{\rm supra}$ is 
%%the threshold that determines whether the transient, differentially 
%%rotating state that appears immediately after merger will transition to a 
%%long-lived NS ($M_{\rm remnant} \leq M_{\rm supra}$) or will eventually 
%%collapse to a BH ($M_{\rm remnant} > M_{\rm supra}$) once differential
%%rotation is removed from the remnant (which typically happens on a secular 
%%timescale~\cite{Shapiro_2000, Ciolfi:2019fie, Ciolfi:2020cpf, Shibata:2021bbj}). 
%%\dr{I think that this separation with the supramassive limit is very
%%  simplistic and should be avoided, see 1803.10865.}
%%\bs{[Agreed. We should at most say that given the TOV/Msupra masses of the isolated 
%%cold equilibria, we EXPECT that...]}\ple{Fixed}
With the exception of models which employ EOS DD2, the total system mass for the 
cases considered in our work falls above $M_{\rm supra}$.
Consistent with this picture, all simulations that employ the DD2 EOS do not 
produce a BH within the end of the simulation. 
Simulations employing other EOS models may form BHs in the 
final state, depending on how the postmerger evolution proceeds. 
%%\dr{BH formation or 
%%not will also depend on how the postmerger evolution proceeds.}\ple{Fixed}

We consider a total of 8 base simulations across the 4 EOS 
models and 2 different mass ratios. We also consider these simulations at lower grid 
resolutions and a subset of them at higher grid resolutions, resulting in a 
total of 20 simulations. 
We construct initial data using the {\tt Lorene} spectral 
solver~\cite{LORENE} for BNS systems using cold 
($T\approx \SI{0.01}{MeV}$), $\beta$-equilibrated slices of the EOS models 
discussed above.
Our initial configurations consist of irrotational binaries with initial 
center-of-mass separations of $\SI{45}{km}$. 
We consider a fixed binary mass of 
$M = \SI{2.7}{M_\odot}$ for all cases,
%\dr{The table says 2.7, and I think it is 2.7 and not 2.68} \ple{Fixed}
%%\dr{I think that
%%only the mass at infinity $M_1 + M_2$ is fixed, the ADM mass will not be
%%the same for all binaries, because it depends on the EOS}\ple{Fixed.}
and take the mass 
ratio to be 
%%\bs{[would put following equation in the text:] $q=M_1/M_2\geq1$,}\ple{Fixed.}
%%\fz{Now I guess the following should be removed}\ple{It helps to have this to clarify that we use this convention for $q$}
\begin{equation}\label{eq:mass_ratio}
q=M_1/M_2 \geq 1,
\end{equation}
where $M_i$ is the mass of a TOV star with the same baryonic mass as 
star $i$ in the binary and the labels $i=1(2)$ correspond to the more 
(less) massive star in the configuration. In 
Tab.~\ref{tab:configurations} we list relevant properties for our set 
of initial data.
%---------------------------------------------------------------
\subsection{Grid setup}
%---------------------------------------------------------------
We consider simulations at 2 grid resolutions, which we label low resolution 
(LR) and standard resolution (SR).
%%\dr{If am not
%%mistake, we have not completed any HR simulations. I think it is
%%misleading to list these simulations.} \ple{Fixed}
The outer 
boundaries of our grid extend to $\sim \pm \SI{1512}{km}$ along 
the $x$- and $y$-directions and to $\sim \SI{1512}{km}$ along the 
$z$-direction; we employ reflection symmetry about the $xy$-plane.
Our solution grid uses 3 sets of boxes nested within the outermost 
boundaries with 
each box employing 7 levels of adaptive mesh refinement (AMR), for which
we use the {\tt Carpet} AMR driver within the 
{\tt EinsteinToolkit}~\cite{Loffler2012ETK}. We place the center of one 
set of boxes at the origin of the solution grid, near the epicenter of the 
merger. The two other sets of boxes are used to track the centers-of-mass 
for each star. The half-side length of the smallest nested boxes extend to 
$r=\SI{14.8}{km}$, such that they fully cover the entirety of each 
NS. The half-side length of the remaining coarser boxes extend to 
$r=\eta~\si{km}$, where
$\eta \in (29.5,59.1,118.2,236.1,443.1)$.
The finest-level grid spacing is 
%\fz{The following are also in [km], right?}\ple{Fixed.}
$\Delta x_{\rm fin}^{\rm LR} \approx \SI{0.25}{km}$ and 
%\dr{LR is 0.25~km}\ple{Fixed throughout}
$\Delta x_{\rm fin}^{\rm SR} \approx \SI{0.185}{km}$
%$\Delta x_{\rm fin}^{\rm HR} \approx \SI{0.0925}{km}$ 
for the LR and SR cases, respectively.
In Tab.~\ref{tab:configurations} we list the 
grid resolutions considered for each configuration in our study.

%===============================================================
\section{Results}\label{sec:results}
%===============================================================
%%\bs{[I'd move subsections in this order: dynamics overview, out-of-equi, ejecta, nu lum, nucleo.]}
%%\fz{Agreed.}
%%\ple{I re-organized things and commented out the out-of-equilibrium section because 
%%it is in the letter.}

\begin{figure*}[htb]
\includegraphics[width=\textwidth]{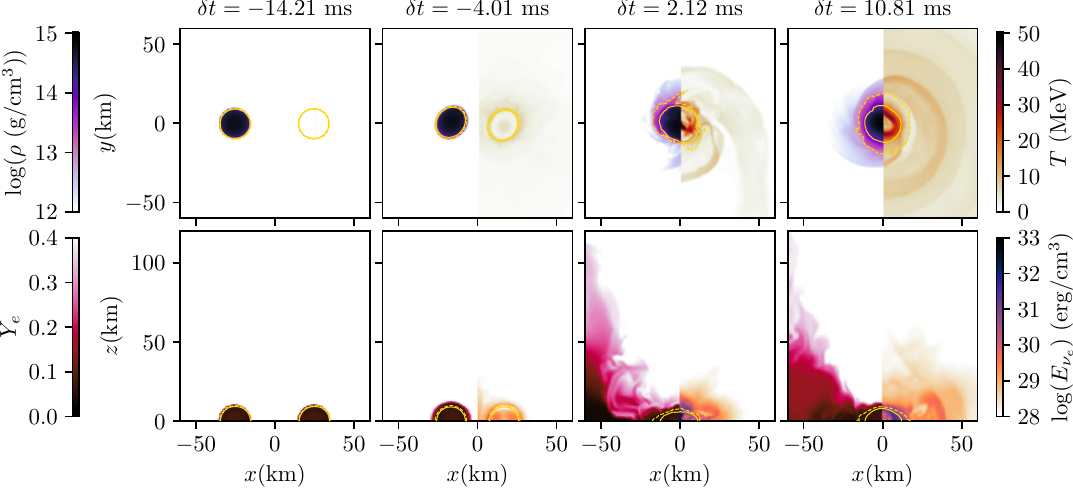}\\
\centering
\caption{
\textit{Top panel:} Equatorial snapshots of the SR \model{DD2}{1} simulation. 
We depict the rest mass density $\rho$ 
on the left half of each panel and the temperature 
$T$ on the right half of each panel, corresponding to the left and right colorbars, 
respectively. From left to right, we show snapshots at times from merger 
%%\fz{The labels on top do not match with these.}\ple{Fixed}
$\delta t \equiv t - t_{\rm mer} \in (-14.21, -4.01, 2.12, 10.81) \SI{}{ms}$. 
We highlight contours of fixed rest mass 
densities 
%%\fz{it's important to stress that $10^{13}$ is conventionally the interface between
%%NS and disk}\ple{Fixed}
$\rho=\SI{e13}{\g\per\cm\cubed}$ and $\rho=\SI{e14}{\g\per\cm\cubed}$ 
using dashed and solid yellow lines, respectively 
(where $\rho=\SI{e13}{\g\per\cm\cubed}$ is conventionally taken to be the interface 
between disk and remnant NS). 
\textit{Bottom panel:} Meridional snapshots for the same simulation depicted 
in the top panel. The left and right half of each snapshot depict the electron 
fraction $Y_{\rm e}$ (corresponding to the left colorbar) and neutrino energy density 
in the lab frame (corresponding to the right colorbar), respectively.
%%\fz{Is the third time fram correct? It seems way more into the post-merger than 0.02 ms..}\ple{the time of merger is identified using GWs and does not necessarily 
%%line up with the time of first contact between stars}
%%\dr{We should use the advanced time for the merger time, see {\tt runs-thc-ba/analysis/modules/utils.py}.} \ple{Fixed}
%%\fz{I would restict z in $[0,25]$ km and make the color bars thicker}\ple{I prefer 
%%this style because it keeps the same total distance in x and y. This thickness of 
%%colorbars allows the color to be seen while leaving more space for the snapshots.}
%%\dr{We could normalize the density by nuclear density, or use CGS.}
%%\dr{There is something strange in the second panel at the bottom.}
%%\ple{Fixed. It was low density stuff which is now masked}
}
\label{fig:contours_density_temp}
\end{figure*}

In the following we highlight the key results from our simulations. We 
focus on the following results: 
%%\fz{Maybe use A,B,C,D,E for this list,
%%to refer to the subsections?}\ple{Fixed} 
(A) we detail the general merger dynamics 
and outcomes, as well as the gravitational radiation extracted from our simulations;
%OOE
%(B) we discuss the microphysical, thermodynamic conditions of the matter in the 
%context of local weak equilibrium; 
(B) we discuss the properties of ejecta; (C) we discuss the general neutrino dynamics and report on neutrino luminosity and 
energetics; (D) we detail the $r$-process
nucleosynthetic yields that arise from our simulations and report on 
approximate kilonova lightcurves. Where relevant, we highlight how our 
improved neutrino treatment plays a role on our results.

%---------------------------------------------------------------
\subsection{Merger dynamics and gravitational waves}\label{subsec:results_dynamics}
%---------------------------------------------------------------

\begin{table}[htb]\label{tab:sim_summary}
\centering
\caption{Summary of key results from our standard resolution (SR) 
simulations. For each simulation we list the EOS, mass ratio $q$, 
approximate time of collapse (after the merger) $\delta t_{\rm coll}$ 
estimated by the time at which the minimum lapse drops below 
$\alpha_{\rm min} < 0.2$, final simulation time $\delta t_{\rm end}$, and frequency 
corresponding to peak 
post-merger GW emission $f_{\rm peak}^{2,2}$. All times are given in ms after the 
merger.
%%\dr{I would give all time in ms after merger. The time of merger from
%%the beginning of the simulation does not really have a physical meaning.}
%%\fz{Agreed, I would remove entirely $t_mer$ column and add $t_{end}$}\ple{Fixed}
}
\begin{ruledtabular}
\begin{tabular}{l | c | c c | c c c }
EOS  & $q$ & $\delta t_{\rm coll} (\SI{}{ms})$ & $\delta t_{\rm end}(\SI{}{ms})$ & $f^{2,2}_{\rm peak} (\SI{}{kHz})$\\ \hline
BLh  & 1.0 & --    & 21.75 & 2.99 \\ \hline
BLh  & 1.2 & 3.29  & 3.82 & 2.94\\ \hline
DD2  & 1.0 & --    & 40.11 & 2.32 \\ \hline
DD2  & 1.2 & --    & 31.20 & 2.36\\ \hline
SFHo & 1.0 & 1.76  & 6.93 & 3.49 \\ \hline
SFHo & 1.2 & 1.87  & 2.33 & 3.21\\ \hline
SLy  & 1.0 & 0.73  & 6.09 & 2.65 \\ \hline
SLy  & 1.2 & 19.18 & 19.55 & 3.38\\
%SRO*  & 1.0   & 2.57472 & 2.84262 \\ \hline
\end{tabular}
\end{ruledtabular}
\end{table}

%%\begin{figure*}[htb]
%%\centering
%%\includegraphics{./strain22_t_f.pdf}
%%\caption{Gravitational waves for a set of representative simulations in 
%%our work. In the top panel we depict the GW strain amplitude for the 
%%dominant $l=2, m=2$ mode in the case of equal (left panel) and unequal
%%(right panel) mass ratios. In the bottom panel we depict the GW signal in 
%%as a function of frequency in the case of equal (left panel) and unequal
%%(right panel) mass ratios. We show results corresponding to 
%%the BLh, DD2, SFHo, SLy, and SLy4 EOS models using solid maroon, dashed 
%%magenta, dotted, purple, dash-dotted blue, and dash-dot-dotted green 
%%lines, respectively. All results depicted are for standard resolution 
%%(SR) simulations.
%%}
%%\label{fig:gw_sr}
%%\end{figure*}

We begin the discussion of our main results with an overview of the general merger 
dynamics observed in our simulations. Specifically, we discuss quantities that 
represent key physics of different stages of the merger, and list some of these 
quantities in Tab.~\ref{tab:sim_summary}. 
The
merger dynamics observed in our simulations cover three 
generic 
scenarios: (1) in some cases we observe relatively short-lived post-merger 
remnants, 
with remnant survival times as short as $t\sim \SI{3}{ms}$ as is the case for 
simulations employing the SFHo EOS; (2) in other cases we observe 
%\fz{I think it
%is a bit ambiguous to call "long-lived" something that collapses at 20 ms. For instance, in
%https://arxiv.org/abs/2111.13005 this case is called "delayed collapse", and long-lived are
%the ones for which collapse is never observed until the end of the simulation. Whatever the choice, be consistent from now on.}\ple{Switched to 
%using ``longer-lived'' throughout manuscript.} 
longer-lived 
%%\fz{also the BH is a remnant object, but here you mean remnant NSs}\ple{Fixed}
remnants NSs that collapse on timescales closer to $t\sim 20-\SI{30}{ms}$, as is the 
case for 
some simulations 
employing the SLy and BLh EOSs; (3) finally, we find cases which produce 
post-merger remnants that do not collapse on the timescales covered by our 
simulations 
(with survival times exceeding $t\sim\SI{50}{ms}$) as is the case for 
simulations 
employing the DD2 EOS. 
In our simulations
we do not observe the prompt-collapse scenario, 
in which a BH is formed immediately at the time of merger.

In the top panel of Fig.~\ref{fig:rhomax_alphmin} we show 
%the maximum rest mass density (top panel) and 
the minimum lapse as a function
of time for equal (left panel) and unequal (right panel) mass ratios. 
Fig.~\ref{fig:rhomax_alphmin} reveals that, depending on mass-ratio, 
different EOS models exemplify scenarios with a short-lived 
(e.g., \model{SFHo}{1}), longer-lived (e.g., \model{SLy}{1.2}), and stable
post-merger remnant on the timescales probed by our simulations 
(e.g., \model{DD2}{1} and \model{DD2}{1.2}), respectively.
%; stages in the simulation 
%where 
%%maximum rest mass density rises above 
%%$\rho_{\rm max} \gtrsim 2.5\rho_{\rm max} (0)$ and 
%the minimum lapse falls below $\alpha_{\rm min} \lesssim 0.2$ are indicative 
%of gravitational collapse to a BH. 
In the lower panel of 
Fig.~\ref{fig:rhomax_alphmin} we show the 
%GW strain as a function of time from merger (top panel) and 
GW spectrum for our simulations. 
%%\fz{I think this text is from an old
%%version, the bands are not there anymore in the fig.}\ple{Fixed}
%%In the GW spectra depicted in 
%%Fig.~\ref{fig:rhomax_alphmin} we also highlight the dominant frequencies 
%%corresponding to the late-inspiral and post-merger stages of the evolution 
%%using green and yellow bands, respectively.
The inspiral signals are very similar in the amplitude and phase between all cases. 
Deviations between cases 
arise mainly in the post-merger stages, where the thermal effects of the EOS 
are expected to manifest. 
%%The late inspiral stages shows significant dephasing between cases, with a 
%%maximum dephasing of X\% between the 
%%DD2 and SFHo cases. 
The high-frequency signal, with $f\geq \SI{2}{kHz}$, is 
dominated by the dynamics of the post-merger remnant, with peak frequencies 
in the range 
$f_{\rm peak} \approx \SI{2.3}{kHz} - f_{\rm peak} \approx \SI{3.5}{kHz}$, 
depending on the EOS. We find that the mass 
ratio does not play a strong role in the value of $f_{\rm peak}$, with shifts 
of at most $\sim 2\%$ across the EOS models considered.

In the top panel Fig.~\ref{fig:contours_density_temp} we show equatorial 
snapshots of the rest mass density $\rho$ and temperature $T$ at different 
stages of the merger using the left and half right of each panel, 
respectively. We focus on results for model \model{DD2}{1} at SR. During the 
inspiral all simulations behave qualitatively similar, with 
negligible oscillations 
%%\fz{Now $\rho_\text{max}$ is not plotted anymore. 
%%I think it should be, though. Any reason why it was removed?}\ple{I felt that the 
%%discussion surrounding $\rho_{\rm max}$ was redundant with the information carried 
%%in $\alpha_{\rm min}$.}
in $\rho_{\rm max}$ and $\alpha_{\rm min}$ (as suggested by the top panel of 
Fig.~\ref{fig:rhomax_alphmin}) and minimal heating (as suggested by the right 
half of the two leftmost frames in Fig.~\ref{fig:contours_density_temp}). 
The first 
significant heating happens at a time near merger, when the two stellar cores 
begin to touch (as depicted in the third-from-left frame in the top panel of 
Fig.~\ref{fig:contours_density_temp} corresponding to $t\approx\SI{2}{ms}$ after the 
merger). 
Some simulations produce a longer-lived remnant massive neutron star 
(RMNS). In these cases, as exemplified by the top 
right panel of Fig.~\ref{fig:contours_density_temp}, the RMNS typically develops with 
a warm core of $T\sim\SI{10}{MeV}$ which is surrounded by an envelope of hotter 
material of $T\sim 30-\SI{40}{MeV}$. This temperature profile
%%\fzedit{s}{} \ple{I am 
%%discussing the singular case depicted in Fig.~\ref{fig:contours_density_temp} here} 
is maintained 
over the lifetime of the remnant as it settles toward an equilibrium state on 
dynamical timescales. 
%%In other cases, 
%%\fz{Which cases? Does it make sense to show a comparison
%%plot? Does anything change when changing $q$?} \ple{The focus of the work was the 
%%effects of neutrinos and neutrinos do not play a role in the dynamics highlighted 
%%here, so I kept the discussion short and focused on one case. This is all pretty 
%%standard in BNS merger simulations (neutrinos do not appear to change any of the 
%%dynamics discussed here) and as such I just mentioned it and referred to the 
%%literature.} 
%%however, the 
%%RMNS develops with a hot core of $T\gtrsim\SI{40}{MeV}$ and decreases in 
%%temperature near the outer layers of the remnant. Such differences in thermal 
%%profiles of longer-lived RMNS may be due to differences in the thermal 
%%properties between EOS models~\cite{Bauswein:2010dn, raithel2021realistic, Raithel:2022nab, Fields:2023bhs}.
%%\dr{I do not remember seeing cases with maximum temperature in the core.}

During the inspiral we find $Y_{\rm e}$ values that reflect the neutron rich 
conditions consistent with cold neutrino-less beta equilibrium, as expected. After 
the merger the high density 
regions comprising the remnant (highlighted with yellow contours in 
Fig.~\ref{fig:contours_density_temp}) remain very neutron rich. The disk 
surrounding the remnant remains neutron rich (with $Y_{\rm e} \lesssim 0.25$) along 
the orbital plane well after the merger, as depicted in the bottom right panel of 
Fig.~\ref{fig:contours_density_temp}. As the angle with the equatorial plane 
increases, so does the typical $Y_{\rm e}$ value, with the region within 
approximately $30\deg$ from the polar axis being comprised of proton rich material 
(with $Y_{\rm e} \geq 0.4$). As discussed in Sec.~\ref{subsec:results_ejecta}, the 
amount of ejected material with $Y_{\rm e} \geq 0.4$ may be significantly enhanced 
when M1 neutrino transport is considered, relative to simulations that use lower-
order neutrino transport schemes~\cite{Zappa:2022rpd}. 

\subsection{Ejecta}\label{subsec:results_ejecta}
%---------------------------------------------------------------

\begin{table*}[htb]\label{tab:ejecta_properties}
\centering
\caption{Summary of key ejecta properties for the SR simulations in our 
study. For each simulation we list the EOS, mass ratio $q$, time after the merger 
at which an AH first forms in the simulation $\delta t_{\rm AH}$,
total ejected mass $M_{\rm ej, tot}$,
ejecta kinetic energy $E_{\rm kin}$ and several mass-averaged quantities, 
including the asymptotic speed $\langle v_{\infty} \rangle$, electron fraction 
$\langle Y_{\rm e} \rangle$, specific entropy $\langle s \rangle$, and temperature 
$\langle T \rangle$. We also show the total amount of ``fast'' ejecta 
(with $v_\infty \geq 0.6$)
$M_{\rm ej}^{v\geq 0.6}$, total amount of ``proton-rich'' ejecta 
(with $Y_{\rm e} \geq 0.4$) $M_{\rm ej}^{Y_{\rm e}\geq 0.4}$, and total amount of 
``shocked'' ejecta (with $s \geq 150$ $k_{\rm B}/{\rm baryon}$)
$M_{\rm ej}^{s\geq 150}$.
}
\begin{ruledtabular}
\begin{tabular}{l c | c c c | c c c c | c c c c c }
EOS  & $q$ & $\delta t_{\rm AH}$ & $M_{\rm ej, tot}$ & $E_{\rm kin}$ & 
$\langle v_{\infty} \rangle$ & 
$\langle Y_{\rm e} \rangle$ & $\langle s \rangle$ & $\langle T \rangle$ & 
$M_{\rm ej}^{v \geq 0.6} $ & $M_{\rm ej}^{Y_{\rm e} \geq 0.4}$ & 
$M_{\rm ej}^{s \geq 150}$\\ 
 & & (ms) & $(10^{-2} M_\odot)$ & $(10^{50} \SI{}{erg})$ &  & & 
 $(k_{\rm B}/{\rm baryon})$ & (MeV) &  $(10^{-2} M_\odot)$ & $(10^{-2} M_\odot)$ &
 $(10^{-2} M_\odot)$\\
\hline
BLh & 1.0 & 21.745 & 0.188 & 0.605 & 0.145 & 0.309 & 23.747 & 0.418 & $\SI{5.949e-04}{}$ & 0.061 & $\SI{5.603e-04}{}$ \\
DD2 & 1.0 & 40.096 & 0.508 & 0.673 & 0.079 & 0.308 & 19.182 & 0.509 & $\SI{3.504e-08}{}$ & 0.144 & $\SI{2.277e-04}{}$ \\
SFHo & 1.0 & 1.762 & 0.819 & 6.807 & 0.287 & 0.283 & 16.761 & 0.697 & $\SI{1.407e-02}{}$ & 0.017 & $\SI{9.041e-04}{}$ \\
SLy & 1.0 & 0.792 & 0.342 & 3.765 & 0.324 & 0.230 & 16.190 & 0.498 & $\SI{1.993e-02}{}$ & 0.013 & $\SI{1.219e-03}{}$ \\
DD2 & 1.2 & 31.181 & 0.363 & 0.530 & 0.071 & 0.266 & 18.494 & 0.400 & $\SI{1.069e-04}{}$ & 0.068 & $\SI{2.220e-04}{}$ \\
SLy & 1.2 & 19.179 & 0.871 & 3.525 & 0.152 & 0.187 & 13.562 & 0.327 & $\SI{8.315e-03}{}$ & 0.101 & $\SI{8.168e-04}{}$ \\
\end{tabular}
\end{ruledtabular}
\end{table*}
\begin{figure*}[htb]
\centering
\includegraphics[scale=1.0]{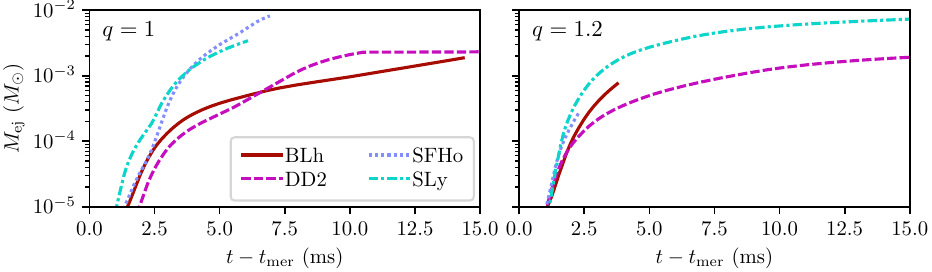}
\caption{Total ejecta mass for the SR simulations considered in this work. In the left (right) panel we show results for the $q=1$ ($q=1.2$) simulations.
}
\label{fig:Mej_sr}
\end{figure*}

\begin{figure*}[htb]
\centering
\includegraphics[scale=1.0]{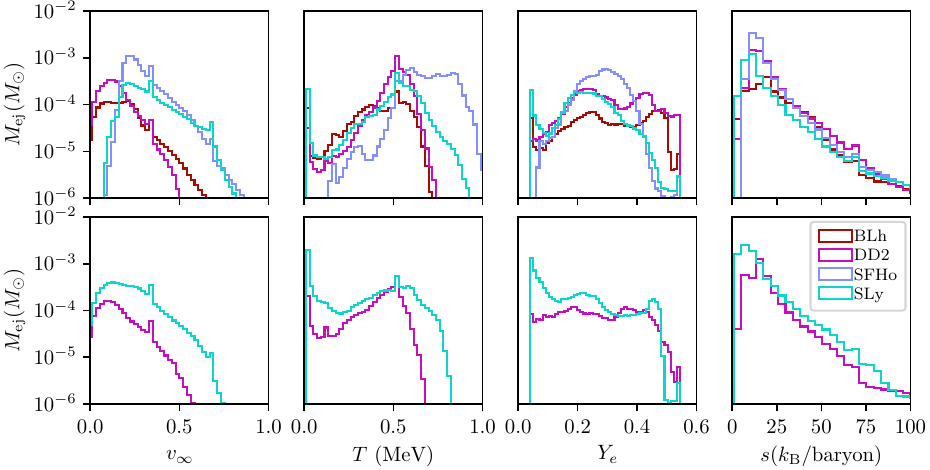}
\caption{Histograms of ejecta properties for simulations of equal ($q=1$, top 
panel) and unequal ($q=1.2$, bottom panel) mass ratios. Specifically, we show the 
total time-averaged ejecta mass for several fluid variables
$M_{\rm ej} = \sum_i m_i$ (where $m_i$ is the mass 
fraction in each variable bin), including the asymptotic velocity $v_{\infty}$, 
temperature
$T$, electron fraction $Y_{\rm e}$, and specific 
entropy $s$. All results depicted are for SR 
simulations which produce a significant amount of ejecta.
%\dr{Are we plotting the density, or the mass in each bin? Meaning is 
%$M_{ej}= \sum M_{i} \Delta Y_{e}$ or $M_{ej} = \sum M_{i}$?}
%\ple{Here we are plotting $M_{ej} = \sum M_{i}$. I've now clarified this in the 
%caption}
%%\dr{What are the colored bands?}\ple{Removed}
%%\bs{[the T-histo is really strange, check?]} \ple{The $T$ histograms were not binned very sensitively. Fixed.}
}
\label{fig:histograms_sr}
\end{figure*}
In Fig.~\ref{fig:Mej_sr} we show the total amount of ejected mass as a function of 
time for the SR simulations. We find that only some cases result in a 
significant amount of ejecta due to the duration of the simulations.
%%\dr{I would maybe be more explicit and say that we exclude simulations
%%A, B, C, as we could not carry them out to sufficiently late times.}\ple{Fixed}
Specifically, 
we exclude models \model{BLh}{1.2} and \model{SFHo}{1.2} from the discussion on 
ejecta because we could not carry 
out the simulations to sufficiently late times.

In Tab.~\ref{tab:ejecta_properties} we summarize the key average ejecta properties 
pertaining to the subset of models which produce significant ejecta. 
For cases with $q=1$ ($q=1.2$), models 
$\text{SFHo}_{q=1}$ ($\text{SLy}_{q=1.2}$) produce the most ejecta, with both cases 
producing close to $\SI{0.01}{M_\odot}$. We note qualitative trends in the 
mass-averaged ejecta properties that potentially reflect the main effects of 
the EOS and the use of M1 neutrino transport. For instance, we note that the 
`stiffer' EOSs we consider (namely, models BLh and DD2) produce ejecta 
with higher $\langle Y_{\rm e} \rangle$, but lower $\langle v_\infty \rangle$. 
Qualitatively, stiffer EOS models (such as BLH and DD2) allow for higher maximum 
remnant masses and lead to longer remnant lifetimes~\cite{Margalit2017}; the 
production of metastable RMNSs
which survive on significantly longer timescales for stiffer EOS models is 
demonstrated by the $\delta t_{\rm AH}$ quantity in 
Tab.~\ref{tab:ejecta_properties}.  
On the other hand, `softer' EOS models such as SFHo and SLy result in more compact 
binary components that undergo more violent collisions at the merger and produce 
stronger shocks~\cite{Radice:2018pdn}. 
The aforementioned trends we observe in $\langle Y_{\rm e} \rangle$ 
and $\langle v_\infty \rangle$ may be attributed to the combined effects of the EOS 
and M1 neutrino transport. Where `softer' EOS models lead to more violent 
shocks at the time of merger and produce higher velocity shocked ejecta, the 
relatively low maximum remnant masses they allow for result in significantly 
shorter neutrino irradiation times in the post-merger which in turn allows the disk 
surrounding the remnant to remain relatively neutron rich. 
On the other hand, `stiffer' EOS models may produce 
lower velocity shocked ejecta while allowing for longer-lived RMNSs that irradiate 
the disk with neutrinos and drive the electron fraction in the system toward higher 
values~\cite{Zappa:2022rpd}.
\begin{figure*}[htb]
\centering
\includegraphics[scale=0.4]{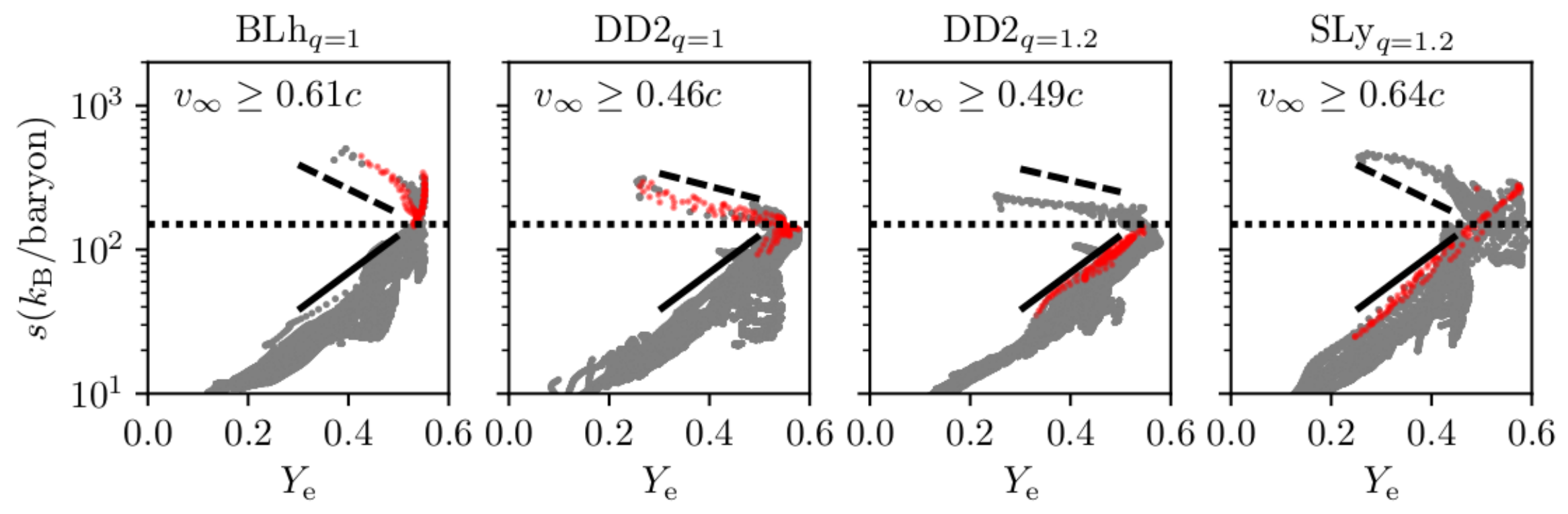}
\caption{Relation between the mass-and-azimuthally-averaged specific entropy $s$ 
and electron fraction $Y_{\rm e}$ for ejecta in four representative SR
simulations in our work. We highlight the ejecta with asymptotic velocity 
$v_\infty$ that falls above the $99^{\rm th}$ percentile of the distribution (with value shown in each panel) using 
red marks.
%%\dr{I would
%%maybe use a different quantity, rather than $v_{\infty, \rm max}$} \ple{Fixed}
For 
reference, we also show approximate fits for potentially disparate components of 
the ejecta using dashed and solid black lines, and  mark the the threshold above 
which we could potentially mark 
shocked ejecta ($s \geq 150\text{ }k_{\rm B}/{\rm baryon}$) with a black dotted 
line.
}
\label{fig:scatter_s_ye}
\end{figure*}

In Fig.~\ref{fig:histograms_sr} we show histograms of relevant ejecta properties 
for SR simulations which produce a significant amount of ejecta. 
In~\cite{Zappa:2022rpd} it was found that the M0 and M1 schemes as implemented in
the {\tt THC} code result in qualitatively similar ejecta. The main difference 
between the predictions of each scheme, in the context of ejecta, is reflected in 
the electron fraction distribution. Specifically, when compared to M0 cases,
simulations employing the M1 scheme may produce significantly more 
proton-rich ejecta. This is reflected in the ejecta 
distribution of $Y_{\rm e}$ for the simulations presented in 
Fig.~\ref{fig:histograms_sr}. In particular, the distributions corresponding to 
models $\text{BLh}_{q=1}$, $\text{DD2}_{q=1}$, $\text{DD2}_{q=1.2}$, and 
$\text{SLy}_{q=1}$ simulations exhibit a 
significant amount of high-$Y_{\rm e}$ ejecta 
(with $Y_{\rm e} \geq 0.4$). 
The relative increase of high-$Y_{\rm e}$ ejecta in M1 simulations is attributed to 
the ``protonization'' of ejecta fluid elements which absorb neutrinos that 
originate in the central object and disk~\cite{Zappa:2022rpd};
neutrino absorption in the ejecta leads 
to a systematic increase of the electron fraction. We note that the aforementioned 
models all produce metastable RMNSs that survive until the end of the simulation, 
which is consistent with the picture of longer-lived RMNSs as neutrino sources
that produce significant proton-rich material.
%%We also note that higher accuracy neutrino transport schemes result in higher 
%%amounts of proton-poor ejecta~\cite{Zappa:2022rpd};
%%the key reason for increased ejecta at higher values 
%%of $Y_{\rm e}$ is neutrino absorption and, as such, schemes that can more 
%%accurately capture the effects of neutrino absorption are expected to produce 
%%more neutron-poor ejecta.

Ejecta profiles extracted from BNS merger simulations may suggest the existence of 
correlations between certain ejecta 
properties~\cite{Nedora:2020qtd, Camilletti:2022jms}. 
Of particular interest is the fast ejecta, which may constitute a dense 
environment through which the jet needs to breakout to power prompt GRB 
emission~\cite{Beloborodov:2018zcy}, may potentially
produce non-thermal emission as it shocks the interstellar medium~\cite{Nakar:2011cw}, 
may power a
short (${\sim}1$ hour) UV transient due to free neutrons
decay~\cite{Metzger:2014yda, Combi:2022nhg, Combi:2023yav}, and may be the origin of the late-time X-ray excess
associated with GW170817~\cite{Hajela:2021faz, Nedora:2021eoj}.
%%\dr{If by fast you mean here $v_\infty \geq 0.6$, then I do not
%%think it can be related to the blue kN. I think the fast ejecta is
%%interesting because 1) it constitutes a dense environment through which
%%the jet needs to breakout to make a GRB; 2) it might potentially
%%produce non-thermal emission as it shocks the ISM; 3) it might power a
%%short (${\sim}1$ hour) UV transient due to free neutrons decay.}\ple{Fixed the discussion to include this and added some references}
If indeed strong 
correlations exist between particular properties of the fast ejecta, this may 
inform modeling efforts targeted at explaining the aforementioned 
phenomena~\cite{Bauswein:2013yna, Ishii:2018yjg, Radice:2018pdn, Dean:2021gpd}.
In the left panel of Fig.~\ref{fig:histograms_sr} we show histograms for the 
asymptotic velocity $v_\infty$. 
High speed ejecta are expected to be 
produced in the violent shocks that arise during the merger~\cite{Nedora:2020qtd}. 
The neutrino scheme 
employed is not expected to play a significant 
role in determining the amount of fast ejecta~\cite{Zappa:2022rpd}, 
so we instead focus on the effects of the EOS 
when discussing the properties of fast ejecta. The increased amount 
of fast ejecta in simulations with relatively soft EOSs (e.g., SFHo and SLy) 
is consistent with the 
picture that such EOS models result in relatively compact binary components which 
undergo relatively violent mergers and in turn produce higher velocity 
ejecta~\cite{Nedora:2021eoj} (see 
Tab.~\ref{tab:ejecta_properties} for reference, where we show the amount of ejecta 
with $v_{\infty} \geq 0.6$).
We note that variability in the amount of fast ejecta which is reflected by the 
tails of the $\langle v_\infty \rangle$ the histograms in 
Fig.~\ref{fig:histograms_sr}. For instance, focusing on 
the top panel of Fig.~\ref{fig:histograms_sr}, models 
$\text{BLh}_{q=1}$ and $\text{DD2}_{q=1}$ show very few ejecta with 
$v_\infty \geq 0.45$ and $v_\infty \geq 0.6$, respectively. On the other hand, 
models $\text{SFHo}_{q=1}$ and $\text{SLy}_{q=1}$ show a significant amount of 
ejecta with $v_\infty \geq 0.6$.
%%\dr{I think we argued along the same lines in Nedora+2020.}\ple{This work is now 
%%cited}

We focus on the relationship between the mass-and-azimuthally-averaged 
specific entropy and electron fraction as an 
example of quantities that show a potential correlation, which we show in 
Fig.~\ref{fig:scatter_s_ye} for a representative subset of our simulations. In 
Fig.~\ref{fig:scatter_s_ye} 
we highlight the component of the ejecta which falls above the $99^{\rm th}$ 
percentile in asymptotic speed (for clarity we show the value of this speed as labels 
in Fig.~\ref{fig:scatter_s_ye})
%%\dr{Can we use the 90th
%%percentile or 99th percentile, instead of $0.9 \times$ the maximum?}\ple{Done, 
%%using 90th percentile}
using red markers and show the 
remainder of the ejecta in gray.
%% \dr{I would color-code the mass of ejecta in each bin, because otherwise
%% we can give the misleading impression that a lot of ejecta has entropy
%% above 100~kb.}\ple{These were made using mass-averaged profiles, so it is hard to 
%% color-code by mass. I have written a statement to emphasize the total amount of 
%% ejecta above ~100 kb is small (in the next paragraph) and now show the total amount 
%% of ejecta with entropy above ~100 kb in Tab.~\ref{tab:ejecta_properties}.}
We also show approximate fits for two potentially 
disparate components of the ejecta using solid and dashed black lines. We find 
that, 
depending on the EOS and mass ratio, it may be possible to identify disparate 
components of the ejecta based on whether it is fast or not. 
For example, for models \model{BLh}{1} and \model{DD2}{1} we find that using the 
aforementioned criterion to 
label fast ejecta results in a unique anti-correlation between $s$ and $Y_{\rm e}$ 
which closely follows the approximate fit represented by the dashed line (as shown 
in the leftmost panel of Fig.~\ref{fig:scatter_s_ye}), whereas the remainder of the 
ejecta approximately follows the trend highlighted by the solid black line. 
However, the trend reflected in the fast ejecta for 
models \model{BLh}{1} and \model{DD2}{1} is not robust 
across different mass ratios or EOS models. For example, the results for models 
\model{DD2}{1.2} and \model{SLy}{1.2} 
%%\dr{Is this $\geq 0.84c$ or $\geq 0.84
%%v_{\max}$? What if we use the 99th percentile of the velocity?}\ple{The figures 
%%now show the 99th percentile}
show that we cannot isolate the fast ejecta 
as following a different trend than the remainder of the ejecta.
We have additionally considered a fixed criterion of $v_\infty \geq 0.6$ to label 
the fast ejecta, but this leads to even higher variability in the potential 
correlations depicted in Fig.~\ref{fig:scatter_s_ye} across different EOS models 
and mass ratios.

We note that for all simulations considered here it may 
be possible to identify a separate component based instead on the specific entropy. 
It is clear from Fig.~\ref{fig:scatter_s_ye} that the component of ejecta with 
$s\gtrsim 150$ $k_{\rm B}/{\rm baryon}$ (highlighted by the horizontal black dotted 
line in Fig.~\ref{fig:scatter_s_ye}) shows an anti-correlation between $s$ and 
$Y_{\rm e}$ for all models. The source of this component of the ejecta is likely 
shocks that 
develop during the merger and shortly after, as this is the source of the highest 
entropy and temperature material we observe. 
%%\dr{I think we need to at least speculate on the reason for this
%%correlation.}\ple{I've now included the speculative reason for these correlations 
%%here.}
The aforementioned relations may also be due to the absorption of neutrinos (or lack 
thereof) for different components of the ejecta. For instance, 
the very fastest ejecta may become diluted before absorbing many neutrinos and 
as such 
does not have its $Y_{\rm e}$ increased by neutrino absorption. 
Since $v_{\infty} \propto s$  for
shocked ejecta (as is the case for models \model{BLh}{1} and \model{DD2}{1}), we may 
find that $s$ and $Y_{\rm e}$ are anti-correlated for this component. On the other 
hand, the slower ejecta may have both $Y_{\rm e}$ and $s$ increased by absorption, 
and may end up with $Y_{\rm e} \propto s$ (which may be similar to what occurs with 
neutrino driven winds~\cite{Nedora:2020hxc}).
We emphasize that the total amount of 
ejecta with $s\gtrsim 150$ $k_{\rm B}/{\rm baryon}$ is very small as shown in 
Tab.~\ref{tab:ejecta_properties} (typically $10^{-6}-\SI{e-5}{M_\odot}$).
%\dr{These numbers do not look right: ${\sim}10^{-3}$ is the total ejecta.} 
%\dr{I would quote the range as $10^{-4}{-}10^{-3}$}\ple{Fixed. I was accidentally 
%quoting the numbers in units of $10^{-2}$ from the table}.
Aside from the 
relations depicted in Fig.~\ref{fig:scatter_s_ye}, we have also considered 
potential correlations among other relevant ejecta properties such as 
between
temperature $T$, rest mass density $\rho$, and flux $F$. Besides those discussed 
above, we find no additional
evidence of trends or correlations in the properties of fast or shocked ejecta 
which are 
robust across EOS models and mass ratios. 
%%\dr{The correlations in
%%Fig.~11 look strong to me} \ple{Rephrased to say no additional clear trends were 
%%seen}
Correlations in the properties of 
fast ejecta, if they exist, may require higher accuracy numerical methods to 
reliably capture and are potentially sensitive to the grid resolution and numerical 
methods used~\cite{Bauswein:2013yna, Ishii:2018yjg, Radice:2018pdn, Dean:2021gpd}.
%% \dr{The way I read this is that the very fastest ejecta becomes diluted
%% before absorbing many neutrinos, since velocity $\propto$ entropy for
%% shocked ejecta, we have the anticorrelation, while the slower ejecta has
%% both $Y_e$ and $s$ increased by absorption, so it ends up with $Y_e
%% \propto s$, like in a neutrino driven wind.}
%% \dr{It would be interesting to do a more detailed nucleosynthesis study
%% as a follow up. The high-entropy, fast expanding ejecta will produce a
%% lot of $He$.}
 %%\ple{Our results do not really show that the 
%% shocked ejecta has $v \propto s$, but I do agree that for slower ejecta absorption may increase both $Y_{\rm e}$ and $s$.}

%---------------------------------------------------------------
\subsection{Neutrino luminosity}\label{subsec:results_neutrinos}
%---------------------------------------------------------------
\begin{figure*}[htb]
\includegraphics[width=\textwidth]{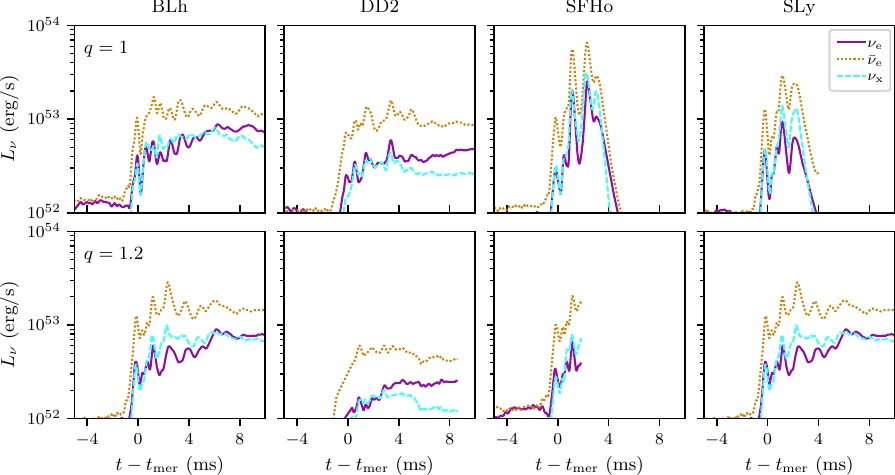}
\centering
\caption{
Neutrino luminosity for the SR simulations in our work. We show results for the 
equal and unequal mass ratio cases in the top and bottom panels, 
respectively.  
%%\dr{The caption is outdated}
%%\dr{What happens at late time for the $q=1$ SLy? Is it real?} \ple{This is after formation of a BH, so probably not. I've removed data after BH formation}
}
\label{fig:neutrino_luminosity}
\end{figure*}

\begin{figure}[htb]
\includegraphics{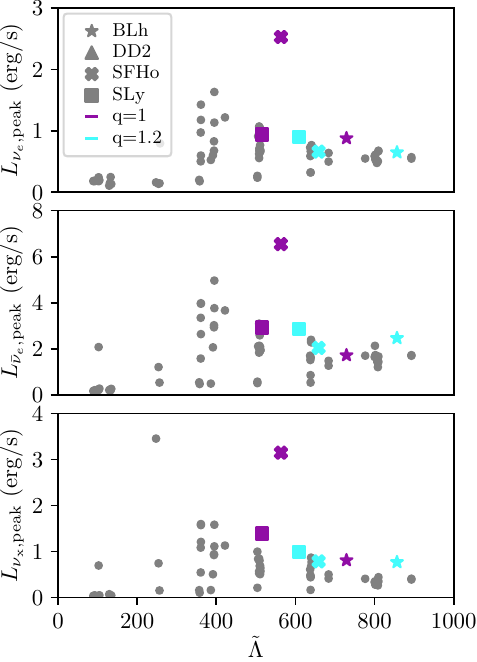}
\centering
\caption{
Peak neutrino luminosity as a function of reduced tidal deformability for the 
SR simulations in this work. Equal (unequal) mass ratio results are shown in purple 
(cyan). For reference, we show the same quantity for the M0 simulations considered 
in~\cite{Cusinato:2021zin} using gray dots.
}
\label{fig:Lnu_Lambda}
\end{figure}

The key new feature in the simulations presented in this work is the use of M1 
neutrino transport, which we briefly reviewed in Sec.~\ref{subsec:M1_equations}. 
In this Section we highlight the main neutrino microphysics 
effects observed in our simulations, including the peak neutrino luminosities 
predicted by our simulations, the production of neutrinos during 
different stages of the merger, and a comparison between the neutrino and GW 
luminosity. Where relevant, we compare our findings to current results in the 
literature which use alternative neutrino transport schemes.

Our simulations show that neutrino production during the inspiral is 
negligible in high-density regions. As suggested by the third frame, top 
panel in Fig.~\ref{fig:contours_density_temp}, the first 
significant shock heating happens at the point of first-contact between the 
binary components, at a time close to $t_{\rm mer}$. Similarly, we find that 
the peak neutrino energy density is reached close to this time (as depicted 
by the third frame, bottom panel in Fig.~\ref{fig:contours_density_temp}). 
In Fig.~\ref{fig:neutrino_luminosity} we show the neutrino luminosity $L_\nu$ 
as a function of time for the SR simulations considered in this work. 
The peak neutrino luminosity is typically reached within 
%%\fz{Is the neutrino flying 
%%time considered here?} \ple{I did not consider that, should I account for the 
%%travel time? Is it fine to just use the retarded time $t-r$?}
%%\dr{You can estimate the travel time as difference in the extraction
%%radii $R_{\rm GW} - R_{\nu}$ and we can correct for that.} \ple{Fixed}
$\SI{4}{ms}$ of the 
merger. We generally find neutrino dynamics and energetics which are compatible 
with findings from throughout the literature and which use alternative neutrino 
transport schemes. For instance, we find general agreement between the peak 
neutrino luminosity and energies predicted by the M1 simulations considered in this 
work and those which 
use a lower-order M0 neutrino treatment~\cite{Cusinato:2021zin} as well as those 
which employ an MC scheme~\cite{Foucart:2022kon} (with peak luminosities on the order 
of $\SI{e53}{erg\per s}$).
%%and peak neutrino energy on the order of
%%$\SI{e50}{erg}$ \dr{$= 10^{55}$~MeV! It cannot be right}\ple{I can't
%%find a source that says the neutrino energy in MC is $\SI{e55}{erg}$})
%%\dr{I converted from erg to MeV. I doubt that this is the average
%%neutrino energy, it is likely off by more than 50 orders of magnitudes.}. 
Moreover, in all simulations we find that the neutrino species follow the same 
order in brightness, with the species corresponding to heavy-lepton flavors 
$\nu_{\rm x}$ being the dimmest, followed by the electron neutrino 
$\nu_{\rm e}$ and finally the electron anti-neutrino $\bar{\nu}_{\rm e}$ being the 
brightest. The general 
qualitative agreement between simulations employing the lower order M0 scheme 
and the simulations presented in this work is encouraging and may be a sign of the 
early convergence of moment based schemes. We refer the reader 
to~\cite{Zappa:2022rpd} for deeper comparisons between the results of the M0 
and M1 schemes as implemented in the {\tt THC} code. 

In Fig.~\ref{fig:Lnu_Lambda} we show the peak neutrino 
luminosity as functions of the tidal deformability of the 
binary for the SR simulations in this work. The reduced tidal deformability is
defined as~\cite{Favata:2004wz}
\begin{equation}\label{eq:reduced_tidal_deformability}
\tilde{\Lambda} \equiv \dfrac{16}{13} \left[ \dfrac{(M_1 + 12 M_2)M_1^4}{(M_1 + M_2)^5} \Lambda_1 + (1\leftrightarrow 2) \right],
\end{equation}
where we use the same labeling convention as in Eq.~\eqref{eq:mass_ratio} and
the tidal deformability of each binary component is 
$\Lambda_i = 2\kappa_2 R_i^5/(3M_i^5)$,
where $\kappa_2$ is the quadrupolar Love number. 
In Fig.~\ref{fig:Lnu_Lambda} we also depict results for the M0 simulations 
considered in~\cite{Cusinato:2021zin} for reference.
Similar to the trend highlighted in~\cite{Cusinato:2021zin} for M0 simulations, 
we note an apparent anti-correlation between the peak 
neutrino luminosity and reduced tidal deformability. The tidal deformability 
of the system increases as the binary components become less compact. As such, the 
merger of systems with larger tidal deformability is relatively less violent 
and results in weaker shock heating of the material during merger. As 
the shocks produced during the merger are key sites for neutrino production 
(see Fig.~\ref{fig:contours_density_temp} for reference), less violent shocks 
result in lower neutrino luminosities. We note that model \model{SFHo}{1} results 
in significantly higher luminosities than all other cases considered, as shown in 
the top panel of Fig.~\ref{fig:neutrino_luminosity}, which results in the model 
appearing as an outlier in the trends depicted in Fig.~\ref{fig:Lnu_Lambda}.
Nevertheless, the remainder of the models we consider show strong support
for the anti-correlation between the peak neutrino luminosity and tidal 
deformability as originally pointed out in~\cite{Cusinato:2021zin}. 

\begin{figure}[htb]
\includegraphics{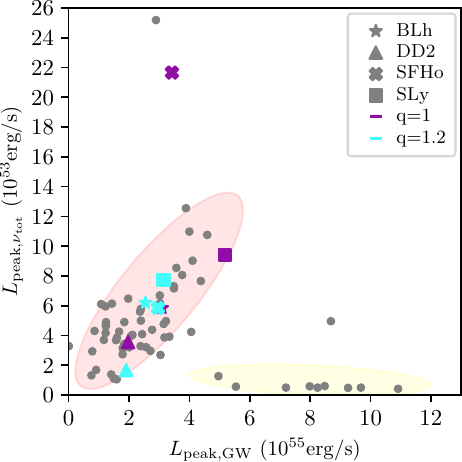}\\
\centering
\caption{
Peak neutrino luminosity $L_{{\rm peak}, \nu_{\rm tot}}$ as a function of peak 
GW luminosity $L_{\rm peak, GW}$ for the SR simulations in our work. Equal 
(unequal) mass ratio results are shown in purple (cyan). For reference we show the 
same quantity for the M0 simulations considered in~\cite{Cusinato:2021zin} using 
gray dots. We highlight potential distinct groups of models based on the lifetime 
of the RMNS produced after the merger; with the yellow and red 
shaded regions roughly corresponding to models that promptly form BHs and models 
that form RMNSs, respectively.
}
\label{fig:Lnua_LGW}
\end{figure}

In Fig.~\ref{fig:Lnua_LGW} we also show the peak neutrino luminosity as a function 
of the peak GW luminosity for the same models considered in 
Fig.~\ref{fig:Lnu_Lambda}. The relationship between peak neutrino and GW 
luminosities originally discussed in~\cite{Cusinato:2021zin} allows for the 
identification of two potential groups of models, 
depending on the lifetime of the RMNS 
formed after the merger. Firstly, for models that promptly collapse to a BH at the time of merger, the peak neutrino luminosity may be 
weakly anti-correlated with the peak GW luminosity; these models mostly reside within 
the yellow shaded region in Fig.~\ref{fig:Lnua_LGW}. Secondly, for models that form 
RMNSs, the peak neutrino 
and GW luminosities may be correlated; these models mostly reside within the red 
shaded region in Fig.~\ref{fig:Lnua_LGW}. In the case of the trends depicted in 
Fig.~\ref{fig:Lnua_LGW} we note that most models considered in this work fall within 
the group of models that show a correlation between $L_{{\rm peak}, \nu_{\rm tot}}$ 
and $L_{{\rm peak, GW}}$ (i.e., within the red shaded region in 
Fig.~\ref{fig:Lnua_LGW}). 
Similar to the potential mechanism behind the 
trends depicted in Fig.~\ref{fig:Lnu_Lambda}, the more violent 
shocks produced during mergers with higher peak GW luminosity results in 
higher neutrino peak luminosities.
We note that all cases considered in our study produce a 
RMNS, albeit with different lifetimes. Variability in the correlation between the 
peak neutrino and GW luminosities may be related to the lifetime of the 
RMNS~\cite{Cusinato:2021zin}, with short-lived remnants that collapse 
within 
$\SI{5}{ms}$ after the merger tending to produce higher neutrino and GW luminosities 
indicative of more violent mergers. However, we find that for the M1 simulations 
considered in our work it is not straightforward to cleanly divide models which 
produce remnants that survive for longer than $\SI{5}{ms}$ after the merger from those 
that do not. For instance, models \model{SFHo}{1} and \model{BLh}{1} produce very 
similar neutrino and GW luminosities, despite producing RMNSs that survive for 
approximately only $\SI{2}{ms}$ and over $\SI{40}{ms}$, respectively. 
%% \dr{The DD2 L\_GW is very large. Maybe it is worth 
%% checking it carefully.}\ple{There was something wrong with the integration at the 
%% extraction radius of r=400, so I used the data for extraction radius of r=500 for 
%% that one instead. I've confirmed that using r=500 does not change anything for the 
%% other models.}
We also note that model \model{SFHo}{1}, which stands out as a 
potential outlier in the trends depicted in Fig.~\ref{fig:Lnu_Lambda}, also stands 
out as a potential outlier in the trends depicted in Fig.~\ref{fig:Lnua_LGW} along 
with one other model which employed M0 neutrino transport.

%---------------------------------------------------------------
\subsection{Nucleosynthesis and kilonova signals}
\label{subsec:results_nucleosynthesis}
%---------------------------------------------------------------
\begin{figure*}[htb]
\centering
\includegraphics{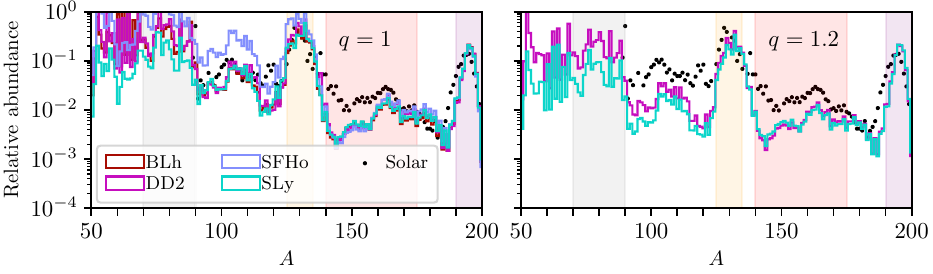}
\caption{Relative abundance of $r$-process elements calculated 
using average ejecta properties extracted from our simulations. 
We normalize such that all models produce the same total abundance for elements 
with 
atomic number $A \geq 170$, which results in similar amounts of third-peak 
$r$-process 
elements in all cases (highlighted using the rightmost shaded region).
We focus on 4 representative models and show the resulting nucleosynthesis 
abundance 
using color lines. For reference, we show the relative abundance for Solar elements 
%%\dr{should this be Solar elements (capitalized)?}\ple{Fixed}
with black dots, using the same normalization.
}
\label{fig:nucleo}
\end{figure*}

\begin{figure*}[htb]
\centering
\includegraphics[scale=0.95]{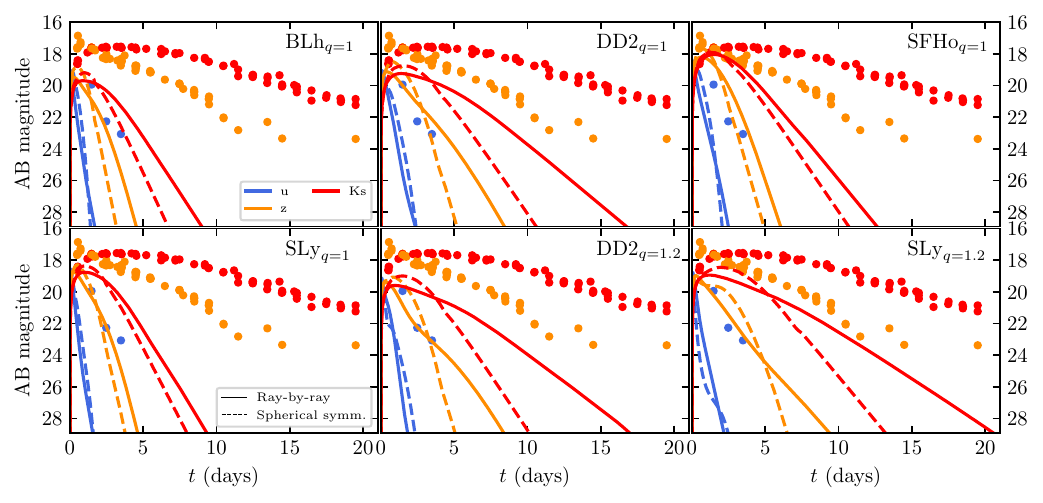}
\caption{Absolute bolometric (AB) magnitude associated with synthetic kilonova 
lightcurves for 6 representative models. 
We show three representative bands using different color lines. 
We compare synthetic light curves computed by assuming spherical symmetry (dashed 
lines) to light curves computed by assuming axisymmetry (solid lines) with a 
ray-by-ray procedure.
For reference, we also show the magnitudes measured 
for AT2017gfo ~\cite{2017ApJ...848L..19C, Cowperthwaite:2017dyu} using dots.
%%\bs{[separate the two plots in two figs]}\ple{Fixed}
}
\label{fig:KN}
\end{figure*}
As discussed in 
Sec.~\ref{subsec:results_ejecta}, a key effect of M1 neutrino transport is the 
capture 
%%\dr{I think that here is where MC does much better than M1 and M1 struggles. I would not use ``accurate''.}\ple{Fixed}
of neutrino absorption which may lead to the significant 
protonization of ejecta (i.e., a significant increase in the amount of 
ejected matter with $Y_{\rm e} \geq 0.4$); this effect is most pronounced when 
comparing the ejecta distributions in $Y_{\rm e}$, 
as shown in Fig.~\ref{fig:histograms_sr}, between 
simulations that lead to longer-lived RMNSs and those that do not. In 
Fig.~\ref{fig:nucleo} we show the relative abundance of elements predicted by 
several SR simulations in our work, along with solar abundances for reference (we 
also highlight the regions roughly corresponding to the first, second, lanthanide, 
and third $r$-process peaks using gray, orange, red, and purple bands, respectively). 
We normalize all abundances such that the total amount of material above $A=170$ is 
equivalent among models, which results in similar abundances in the third $r$-process 
peak elements. All models considered produce second and third peak abundances which 
are consistent with the solar pattern.
%%\dr{The lanthanides peak is very sensitive to nuclear physics
%%uncertainties, so I would not read much into it.}\ple{Fixed}
Among the set of models depicted in
Fig.~\ref{fig:nucleo} we consider cases which result in both short and longer-lived 
RMNSs. Although these differences in post-merger evolution are reflected in the 
$Y_{\rm e}$ distribution of the ejecta, they do not appear to significantly affect 
the nucleosynthetic yield of $r$-process elements. For instance, all models predict 
similar abundances for the second, third, and lanthanide peaks despite 
significantly 
different remnant lifetimes and neutrino irradiation times. Moreover, models 
\model{SFHo}{1} and \model{SLy}{1} (which result in short-lived RMNSs) 
predict larger and smaller first-peak 
abundances, respectively, when compared to models that result in longer-lived RMNS. 
In brief, we do not find clear trends or correlations between the nucleosynthesis 
yields and the RMNS lifetime (and thereby neutrino irradiation time).

\begin{figure}[htb]
\includegraphics{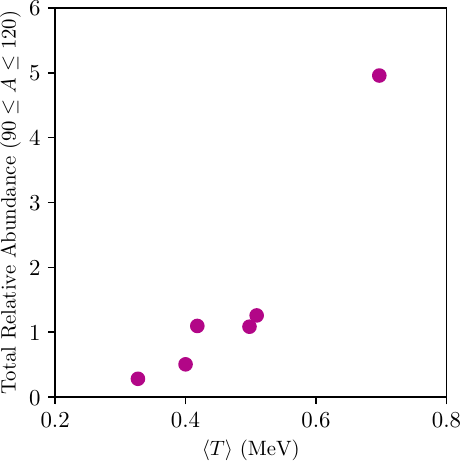}\\
\centering
\caption{
Total abundance of nucleosynthetic yield (relative to the abundance of elements with 
$A \geq 170$, which we normalize for all models) for elements with 
$90 \leq A \leq 120$ as a function of mass-and-time-averaged ejecta temperature (at a 
fixed radius). Each datum corresponds to a single SR simulation from our work.
}
\label{fig:relabund_T}
\end{figure}

A potential 
trend which we \emph{do} note is between the mass-and-time-averaged ejecta temperature (at a 
fixed radius from the source)
%%\dr{I am guessing at a fixed radius, since the temperature is not constant as the 
%%ejecta expand} \ple{This is now noted} 
and the abundance of elements between the first and second $r$-process peaks (i.e., 
with $90 \lesssim A \lesssim 120$). For instance, we note that model \model{SFHo}{1} 
(\model{SLy}{1.2}) produces the largest (smallest) relative abundance of elements 
between the first and second peaks. All other models roughly follow a trend that 
suggests the larger $\langle T \rangle$, the larger the relative abundance of 
elements with $90 \lesssim A \lesssim 120$. In Fig.~\ref{fig:relabund_T} we show the 
total relative abundance of elements with $90 \leq A \leq 120$ (in other words, the 
sum of the relative abundances depicted in Fig.~\ref{fig:nucleo} with 
$90 \leq A \leq 120$) for each model depicted in Fig.~\ref{fig:nucleo}. We see a 
clear correlation between the relative abundance of elements with 
$90 \leq A \leq 120$ and $\langle T \rangle$.
We note that the typical temperatures of the ejecta for the models we 
consider is close to the
threshold for nuclear statistical equilibrium (NSE) 
freeze-out of $T\approx \SI{5}{GK} \approx \SI{0.4}{MeV}$, above which the 
temperature-sensitive 
photo-disintegration reaction cross-sections (which are important in determining 
the abundance of elements with with $90 \lesssim A \lesssim 120$) are expected to 
be large~\cite{Perego:2021dpw}. The sensitivity of the abundance of elements with 
$90 \lesssim A \lesssim 120$ to the average ejecta temperature may be reflective of 
the sensitivity of the photo-disintegration cross-section to temperature.
We emphasize that $\langle T \rangle $ is the average temperature at a fixed radius, 
and that the temperature is expected to change as the ejecta expands.
An additional potential caveat of the preceding discussion is that
%%temperatures extracted from the EOS models considered in this work may not be 
%%very reliable below $T \lesssim \SI{1}{MeV}$~\cite{Hammond:2021vtv}
%%\dr{I think that the EOS is fine for the ejecta, it is radiation-pressure
%%dominated, which is very well known. We might have problems with the surface of
%%the stars in the inspiral, but this is a different issue.}\ple{Comment removed}.
ejecta properties may depend sensitively on the grid resolution considered (see 
App.~\ref{app:convergence} for additional detail).
%%\dr{We have multiple resolutions, so we can comment on this directly, without 
%%referring to Zappa}\ple{Fixed}.
Nevertheless, it is interesting to 
note potential correlations between average ejecta properties such as 
$\langle T \rangle$ and the nucleosynthetic patterns. We leave an investigation of 
the robustness of the potential correlation between $\langle T \rangle$ and the 
abundance of elements with $90 \lesssim A \lesssim 120$, along with other potential 
trends, to future work.
%%\dr{I am guessing that this is because $T$ correlates with $s$ and $s$ with $Y_e$?}
%%\ple{I am not sure, here I am discussing the mass-averaged temperature 
%%$\langle T \rangle$ which does not correlate with $\langle s \rangle$}
%%\dr{For radiation-dominated EOS, which should be relevant for the
%%ejecta, $T$ and $s$ are directly related.}
%%\dr{We need to make it clear that this is temperature at a fixed radius.
%%Temperature is not constant as the ejecta expands. I am wondering if
%%this is actually the best variable to look at.}
%%\ple{I've emphasized that this is $\langle T \rangle$ at a fixed radius. 
%%Please let me know if I should remove this discussion if there is not significant 
%%evidence for it.}
%%\dr{It would be good to have a scatter plot, because we do not provide
%%much evidence for it.} \ple{A scatter plot is now included in Fig.~\ref{fig:relabund_T}}

In Fig.~\ref{fig:KN} we show the absolute bolometric (AB)
magnitudes for kilonova signals associated with a subset of simulations in 
our work; for reference we also show the AB magnitude for the KN signal AT2017gfo. 
We compute light curves in two ways: assuming spherical symmetry and 
assuming axisymmetry. In the latter case ray-by-ray independent evolutions 
are performed for a discrete number of polar angles and the results obtained 
for each angle are then combined together to obtain the total AB magnitude.
We again focus on a subset of representative models to 
understand the effects of the EOS and the mass ratio. 

We generally find KN lightcurves which are 
significantly dimmer than AT2017gfo data. 
However, discrepancies between the predicted lightcurves of BNS merger simulations 
and the AT2017gfo observation are expected and may be due to the relatively short 
duration simulations we consider which do not account for matter ejected on secular 
timescales or uncertainties stemming from multidimensional effects and viewing 
angle~\cite{Perego:2017wtu}. Moreover, we 
did not target the simulations in this work to reproduce the system associated with 
AT2017gfo. As such, we focus the remainder of the discussion around KN on the 
component produced by the \emph{dynamical} ejecta.

We find that all models result in similar dynamical ejecta KN 
lightcurves, typically peaking between 1-2 days in all bands. 
We note that the KN for models \model{SLy}{1} and \model{BLh}{1} (which produce the 
shortest and one of the longest lifetime RMNSs, respectively) both decay within 
approximately 7 days. By comparison, models \model{DD2}{1} and 
\model{SFHo}{1} result in lightcurves which decay after roughly 10 days, despite 
the significant differences in RMNS lifetime between those cases. 
Such differences in  KN lightcurves may not be attributable to the enhanced 
neutrino absorption introduced  by the M1 scheme, but rather to differences in the 
total amount of ejecta that each  model produces. 
With reference to Tab.~\ref{tab:ejecta_properties}, we 
note that models which produce fewer ejecta lead to shorter-duration KN. For 
example, 
model \model{BLh}{1} (\model{SLy}{1.2}) produces the least (most) amount of total 
ejecta and decays the fastest (slowest).
The total amount of ejecta may be 
sensitive to effects such as the  numerical grid resolution 
as we discuss in App.~\ref{app:convergence}, but is not 
significantly affected by the use of the M1 scheme over the M0 
scheme~\cite{Zappa:2022rpd}.
%%\dr{Refer to appendix, compare different resolutions.} \ple{Fixed.}
All cases considered result in KN lightcurves consistent with a dynamical component 
which becomes transparent within a few days. Specifically, models \model{BLh}{1}, 
\model{DD2}{1}, \model{SFHo}{1}, \model{SLy}{1}, \model{BLh}{1.2}, and 
\model{SLy}{1.2} result in dynamical ejecta which becomes transparent after roughly 
1.6, 3.9, 2.5, 2.4, 4.1, and 7.4 days, respectively. This pattern roughly follows the 
evolution time of the KN for each model, with models \model{BLh}{1} 
(\model{SLy}{1.2}) resulting in the shortest (longest) time to transparency and 
slowest (fastest) KN evolution and all other cases resulting in comparable time to 
transparency and evolution times.
We note a general improvement in the light curves calculated with the 
ray-by-ray procedure. Despite the peak of the light curves in this case not being 
brighter than the ones calculated in spherical symmetry, their decay at later times 
($\gtrsim 5$ days after the merger) follows the experimental data more closely. 
This is particularly evident focusing on the Ks band for the models simulated with 
the DD2 EOS, for which we recover an AB magnitude of ${\sim} 24$ 
at 10 days after merger with KNEC computations in axisymmetry,
against ${\sim} 28$ found with KNEC spherical symmetric evolutions.
We expect a further improvement of these results in long-term simulations with
a delayed BH collapse \cite{Zappa:2022rpd}, which include large amount of spiral-wave 
winds \cite{Nedora:2019jhl} and disk winds ejected at secular timescales.

\section{Conclusion}\label{sec:conclusion}
%===============================================================
In this work we have presented an investigation of the combined effects of 
M1 neutrino transport, the EOS model, and the mass ratio in 3D GRHD BNS merger 
simulations. 
%%\dr{But there are papers by Sekiguchi (1502.06660, 1603.01918) and Foucart
%%(1908.00655) that looked at the same effects. It is true that their M1 was not
%%as good as ours, but I would avoid saying that we are the first. }\ple{Fixed.}
The state-of-the-art simulations presented in this work elucidate the role of 
accurate neutrino transport. We find general agreement between 
the predicted neutrino luminosities in M0~\cite{Zappa:2022rpd}, 
M1~\cite{Zappa:2022rpd}, and MC simulations~\cite{Foucart:2022kon}, and note that 
M1 simulations appear to support the potential trends between peak neutrino and GW 
luminosities discussed in~\cite{Cusinato:2021zin}. Although we find general 
qualitative agreement between the simulations presented in this work and those that 
consider alternative neutrino transport schemes such as M0 or MC, we find that there 
are quantitative differences in the prediction of several effects.

Regarding 
ejecta properties, we find that the effects of the EOS and the enhanced neutrino 
absorption introduced by the use of the M1 scheme can conspire to give 
qualitatively different outcomes to what is typically observed using lower-accuracy 
methods such as M0 neutrino transport. 
Along with the details of the viscosity model used (which was 
left fixed for all simulations in this work), the EOS largely determines the 
longevity of the metastable RMNS produced during BNS merger 
simulations~\cite{Margalit2017}. Different RMNS lifetimes can in turn significantly 
impact the neutrino irradiation time, as the main source of neutrino radiation in 
the post-merger stage is the RMNS. When accounting for the accurate neutrino 
re-absorption captured by the M1 scheme, differences in the neutrino irradiation 
time result in significant differences in the average electron fraction of the system. 
Specifically, simulations that produce longer-lived RMNSs lead to relatively high 
average electron fractions when compared to simulations that result in short-lived 
RMNSs, reflective of the protonization effect introduced by neutrino absorption. 
The amount of high $Y_{\rm e}$ ejecta (with $Y_{\rm e} \geq 0.4$) is typically a 
factor of 6-10 times larger (see Tab.~\ref{tab:ejecta_properties} for reference) in simulations that produce RMNSs that survive over the 
duration of the simulations (typically at least $\SI{15}{ms}$ post-merger). Such 
differences in the amount of high $Y_{\rm e}$ material is not reliably captured by 
lower accuracy neutrino transport schemes~\cite{Zappa:2022rpd}, and appears to be a 
novel feature captured by the M1 simulations presented here. We find that the use of an M1 scheme does not 
significantly impact the amount of shocked fast ejecta produced during the merger.
We note that it may be possible to identify disparate components of the ejecta 
based on the asymptotic speed (see Fig.~\ref{fig:scatter_s_ye} for reference). 
However, these trends are not robust across different EOS models or mass ratios. 
%%and we highlight the 
%%difficulty in identifying potential trends between the properties of the fast 
%%ejecta due to uncertainties in the EOS and criterion used to label ejecta as 
%%``fast'' (see Fig.~\ref{fig:scatter_s_ye} for more detail)
%% \dr{I would not say that 
%%this is a major conclusion of the work. Instead, it seems to me that you have 
%%identified clear trends.}\ple{I've fixed the discussion to point out the trends, 
%%while mentioning that it is sensitive to the EOS and mass ratio}. 
Interestingly, across our 
simulations it \emph{is} possible to separate different components of 
the ejecta based on the entropy. In particular high-entropy ejecta, likely stemming 
from the production of shocks, appears to show an anti-correlation between the 
specific entropy and electron fraction. 

We find that the nucleosynthesis
pattern does not reflect the aforementioned enhancement of high $Y_{\rm e}$ 
material introduced by the M1 scheme,
and is potentially more sensitive to the 
the average ejecta temperature $\langle T \rangle$.
%%\dr{I think that the 
%%nucleosynthesis is sensitive to the amount of material with $Y_e > 0.25$}\ple{But 
%%this is not shown from our results, or do you mean that the nucleosynthesis is 
%%sensitive to the amount of material specifically with $0.25 < Y_e <
%%0.4$?} \dr{I think the issue is that M1 shifts high Ye material to
%%higher Ye, but that is not consequential for the nucleosynthesis, which
%%rather depend on the relative amount of Ye > 0.25 and Ye < 0.25 material.}. 
Specifically, we note a 
potential correlation between the relative abundance of elements with 
$90 \lesssim A \lesssim 120$ and $\langle T \rangle$. Finally, we note that all 
models considered lead to qualitatively similar synthetic KN lightcurves consistent 
with \emph{dynamical} ejecta, while showing variation in the evolution time and time 
at which the ejecta become transparent.
%%\dr{The
%%difference in the time at which the ejecta become transparent (the
%%lanthanide curtain effect) is actually quite different for the different
%%models. I think we should emphasize this.}\ple{Fixed}
The decay time of the synthetic KN 
lightcurves we 
calculate appear to be sensitive to the total amount of ejecta produced by each 
model, which is not strongly affected by the neutrino transport 
scheme~\cite{Zappa:2022rpd}.
%%\dr{I would suggest to refocus the kN discussion on the contribution of the
%%dynamical ejecta, rather than the full kN. Here the two interesting outputs of
%%SNEC are 1) the luminosity of the dynamical ejecta (which we show) and 2) the
%%time at which the dynamical ejecta becomes transparent (which we should briefly 
%%discuss).}\ple{I've refocused the KN discussion on the dynamical ejecta and included 
%%the time at which it becomes transparent.}

In this work we have presented several key BNS merger phenomena
where accurate neutrino transport 
may play a role. 
Our aim was to consider a wide variety of 
phenomena which could be impacted by the neutrino transport scheme. The results 
presented in this work leave open many avenues of investigation. For example, 
future work may consider the robustness of the effect of M1 neutrino transport in 
producing enhanced amounts of high-$Y_{\rm e}$ materials for cases that produce 
longer-lived RMNSs, by considering considerably longer post merger evolutions. In 
particular, it would be interesting to consider whether the enhanced protonization 
of the medium introduced by the M1 scheme continues long after the merger, or if 
there is a limit 
to the amount of high-$Y_{\rm e}$ ejecta produced. 
Taking advantage of the reliable neutrino absorption capture by 
the M1 scheme, it would also be interesting to quantify the size and nature of 
emergent bulk viscosity from out-of-equilibrium dynamics~\cite{Espino:2023dei}. 
%A calculation of the size 
%and role of the bulk viscosity using M1 neutrino transport would be the first of 
%its kind in the context of 3D GRHD BNS merger simulations, and would provide 
%crucial information about whether the effects of emergent bulk viscosity should 
%always be considered.
We leave a full investigation of the effects of M1 neutrino transport on longer-lived 
RMNS environments and a calculation of the bulk viscosity which arises during BNS 
mergers, with M1 neutrino transport, to future work.

\begin{figure*}[htb]
\centering
\includegraphics[scale=1.0]{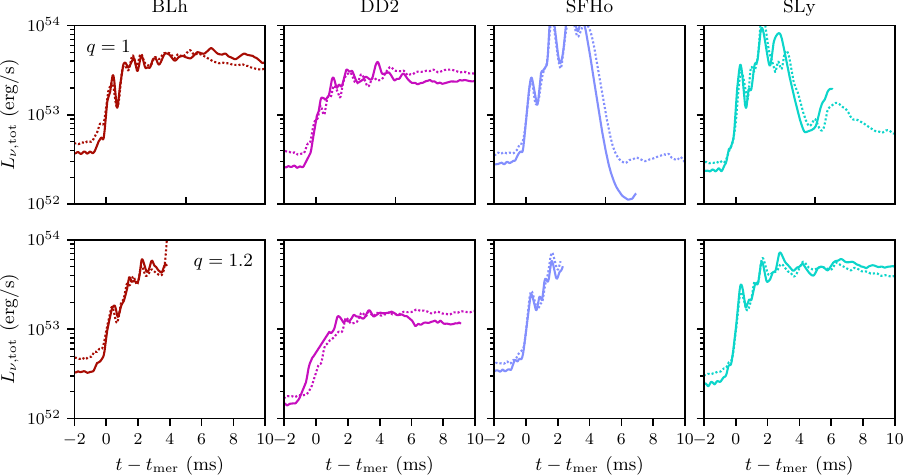}
\caption{Total neutrino luminosity for the LR (dotted lines) and 
SR (solid lines) simulations in our work. The top and bottom panel show results for 
the equal and unequal mass ratio simulations, respectively.
}
\label{fig:neutrino_luminosity_LRSR}
\end{figure*}

%%\begin{figure*}[htb]
%%\centering
%%\includegraphics[scale=1.0]{./Mej_LR.pdf}
%%\caption{Total ejecta mass for the LR simulations considered in this work. In the left (right) panel we show results for the $q=1$ ($q=1.2$) simulations.
%%}
%%\label{fig:Mej_sr}
%%\end{figure*}

\section*{Acknowledgments}
We would like to thank Marco Cusinato for 
sharing the M0 simulation data and useful discussions. We would like to thank Mukul 
Bhattacharya for useful comments and discussions.
PE and RG acknowledge funding from the National Science Foundation under Grant
No. PHY-2020275.
DR acknowledges funding from the U.S. Department of Energy, Office of
Science, Division of Nuclear Physics under Award Number(s) DE-SC0021177,
DE-SC0024388, and from the National Science Foundation under Grants No.
PHY-2011725, PHY-2116686, and AST-2108467.
RG is supported by the Deutsche Forschungsgemeinschaft (DFG) under Grant No.
406116891 within the Research Training Group RTG 2522/1.
FZ acknowledges support from the EU H2020 under ERC Starting
Grant, no.~BinGraSp-714626.  
SB acknowledges support from the EU H2020 under ERC Starting
Grant, no.~BinGraSp-714626, from the EU Horizon under ERC Consolidator Grant,
no. InspiReM-101043372 and from the Deutsche Forschungsgemeinschaft
(DFG) project MEMI number BE 6301/2-1.
Simulations were performed on Bridges2, Expanse (NSF XSEDE allocation
TG-PHY160025), Frontera (NSF LRAC allocation PHY23001), and Perlmutter. 
This research used resources of the National Energy Research Scientific
Computing Center, a DOE Office of Science User Facility supported by the
Office of Science of the U.S.~Department of Energy under Contract
No.~DE-AC02-05CH11231.
The authors acknowledge the Gauss Centre for Supercomputing
e.V. (\url{www.gauss-centre.eu}) for funding this project by providing
computing time on the GCS Supercomputer SuperMUC-NG at LRZ
(allocation {\tt pn36ge} and {\tt pn36jo}).

\appendix

\section{On the potential effects of grid resolution }
\label{app:convergence}
\begin{figure*}[htb]
\centering
\includegraphics[scale=1.0]{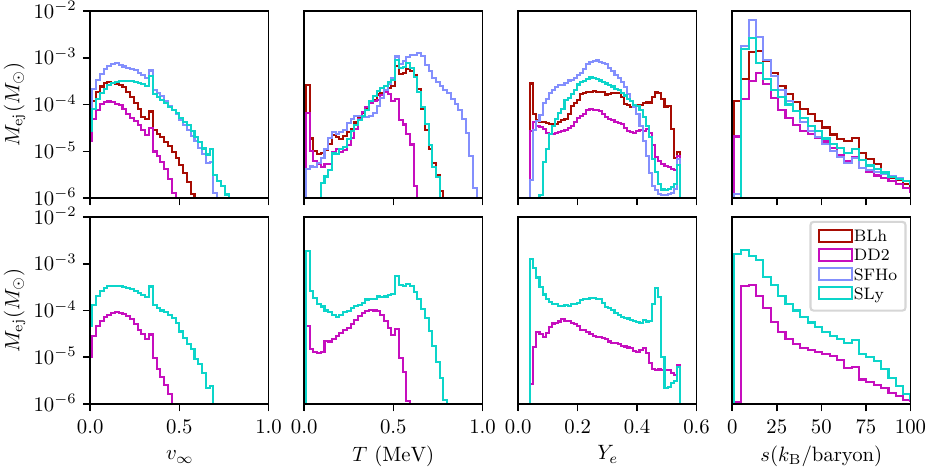}
\caption{Histograms of ejecta properties for LR simulations of equal ($q=1$, top 
panel) and unequal ($q=1.2$, bottom panel) mass ratios. We depict histograms for 
several fluid variables, including the asymptotic velocity $v_{\infty}$, 
temperature $T$, electron fraction $Y_{\rm e}$, and specific 
entropy $s$. We show results corresponding to 
the BLh, DD2, SFHo, and SLy EOS models using solid maroon, dashed 
magenta, blue, and green lines, respectively.
%% \dr{This might be confusing to the reader: why are we talking about LR data, if we are plotting LR data?}\ple{Typo, this data is only for LR}
%% \dr{Maybe plot both SR and LR as in the previous figure?}\ple{This leads to very busy plots}
}
\label{fig:histograms_LR}
\end{figure*}

\begin{table*}[htb]\label{tab:ejecta_properties_LR}
\centering
\caption{Summary of key ejecta properties for the LR simulations in our 
study. 
We list the same quantities considered in Tab.~\ref{tab:ejecta_properties}. 
Additionally, in the rightmost column we show the \emph{absolute} difference
in total ejecta mass between the SR and LR cases 
$\delta M_{\rm ej} \equiv M_{\rm ej, tot}^{\rm SR} - M_{\rm ej, tot}^{\rm LR}$, 
accounting for differences in total simulation time.
}
\begin{ruledtabular}
\begin{tabular}{l c | c c c | c c c c | c c c | c c }
EOS  & $q$ & $\delta t_{\rm AH}$ & $M_{\rm ej, tot}$ & $E_{\rm kin}$ & 
$\langle v_{\infty} \rangle$ & 
$\langle Y_{\rm e} \rangle$ & $\langle s \rangle$ & $\langle T \rangle$ & 
$M_{\rm ej}^{v \geq 0.6} $ & $M_{\rm ej}^{Y_{\rm e} \geq 0.4}$ & 
$M_{\rm ej}^{s \geq 150}$ & $\delta M_{\rm ej}$\\ 
 & & (ms) & $(10^{-2} M_\odot)$ & $(10^{50} \SI{}{erg})$ &  & & 
 $(k_{\rm B}/{\rm baryon})$ & (MeV) &  $(10^{-2} M_\odot)$ & $(10^{-2} M_\odot)$ &
 $(10^{-2} M_\odot)$ & $(10^{-2} M_\odot)$ \\
\hline
BLh & 1.0 & 26.054 & 0.542 & 0.941 & 0.090 & 0.298 & 19.650 & 0.496 & $\SI{1.454e-04}{}$ & 0.140 & $\SI{4.718e-04}{}$  & 0.185 \\
DD2 & 1.0 & 13.539 & 0.166 & 0.336 & 0.110 & 0.259 & 21.274 & 0.402 & $\SI{0.000e+00}{}$ & 0.021 & $\SI{1.682e-04}{}$  & 0.066 \\
SFHo & 1.0 & 2.033 & 1.294 & 5.027 & 0.158 & 0.252 & 14.565 & 0.610 & $\SI{5.342e-03}{}$ & 0.016 & $\SI{1.307e-03}{}$  & 0.019 \\
SLy & 1.0 & 1.072 & 0.609 & 3.723 & 0.215 & 0.272 & 15.398 & 0.529 & $\SI{7.713e-03}{}$ & 0.032 & $\SI{9.772e-04}{}$ & -0.120  \\
DD2 & 1.2 & 10.639 & 0.126 & 0.327 & 0.136 & 0.224 & 18.579 & 0.325 & $\SI{0.000e+00}{}$ & 0.012 & $\SI{1.403e-04}{}$  & -0.017 \\
SLy & 1.2 & 20.679 & 0.789 & 2.959 & 0.144 & 0.189 & 14.432 & 0.322 & $\SI{2.620e-03}{}$ & 0.095 & $\SI{7.820e-04}{}$  & -0.106 \\
\end{tabular}
\end{ruledtabular}
\end{table*}

In order to understand the effects of grid resolution on our results, we consider a 
subset of simulations in our study at different grid resolutions. Specifically, we 
consider all models with lower resolution grids. 
%The {\tt THC\_M1} code is built upon 
%the well-tested {\tt WhiskyTHC} code that demonstrates the expected convergence rates
%in the evolution of relevant fluid quantities~\cite{WHiskyTHC1, 
%WhiskyTHC2} \dr{I would not claim any convergence in this work (unless we have data to demonstrate it)}.
Recently, extensive resolution studies of the {\tt THC\_M1} code have been 
carried out in ~\cite{Radice:2021jtw} and~\cite{Zappa:2022rpd}, and we refer the 
reader to those works for a clearer understanding of the effects of resolution. 
Here we focus on the effects 
of resolution on the key microphysics quantities considered in 
Sec.~\ref{sec:results}, with particular focus on the neutrino luminosities 
and ejecta properties. 

In Fig.~\ref{fig:neutrino_luminosity_LRSR} we show the total neutrino luminosity for 
the LR and SR simulations in our work. We find similar luminosities at both grid 
resolutions, with the LR simulations predicting slightly higher (by typically 20\%) 
luminosities prior to the merger but very similar (with relative difference of 
at most 10\%) after the merger. We also find that the order of brightness between 
neutrino species is the same for LR and SR simulations, and is consistent with 
findings using M0~\cite{Cusinato:2021zin} and MC~\cite{Foucart:2022kon} neutrino 
transport. This suggests that uncertainties associated with low grid resolutions 
are as important as uncertainties associated with using approximate neutrino 
transport schemes, and as such low-resolution results using full neutrino transport 
schemes may not be reliable for the calibration of approximate methods. 
%%\dr{This 
%%last statement should be emphasized: finite-resolution
%%effects are as important as neutrino-transport error (implying low-resolution
%%fancy neutrino transport calculations are not useful).}\ple{Fixed}

In Fig.~\ref{fig:histograms_LR} we show histograms for the LR simulations of 
the ejecta mass in the relevant quantities discussed in 
Sec.~\ref{subsec:results_ejecta}. We find that the protonization effect introduced by 
enhances neutrino irradiation by the RMNS in the M1 scheme is not captured to the 
same extent in the LR simulations as it is in the SR simulations. Although 
we \emph{do} see a relative increase in the amount of ejecta with 
$Y_{\rm e} \geq 0.4$ for LR 
simulations, the enhancement is not as large as we observe for the SR simulations 
discussed in Sec.~\ref{subsec:results_ejecta}. The highest relative increase in the 
amount of high-$Y_{\rm e}$ (with $Y_{\rm e} \geq 0.4$)
%%\dr{Maybe worth repeating that this is Ye > 0.4}\ple{Fixed} 
ejecta when comparing SR simulations with a longer-lived 
RMNS to those with short-lived RMNS is approximately a factor of 12 (specifically 
when comparing the SR \model{DD2}{1} and \model{SFHo}{1} models). On the other hand, 
the highest relative increase we observe for LR simulations is approximately only a 
factor of 9  (when comparing the analogous LR models the LR \model{BLh}{1} and 
\model{SFHo}{1} models). This potential effect of the grid resolution is mostly 
reflected when comparing the LR and SR simulations for model \model{DD2}{1}, which 
predict $\SI{0.021e-2}{M_\odot}$ and $\SI{0.144e-2}{M_\odot}$ of high-$Y_{\rm e}$ 
material, respectively. The relatively lower amounts of high-$Y_{\rm e}$ material for 
the LR simulations considered in our work may also be conflated by having a different 
duration for each simulation.
%%\dr{Can we do a more fair comparison by
%%using the same postmerger time for LR and SR?}\ple{I did this for the things we 
%%quantitatively compared such as the total ejecta mass and average properties, but not 
%%for the qualitative comparisons such as the ejecta distributions. I think we will not 
%%get much more insight for the qualitative comparisons if we do this.} 
It may be the 
case that the longer 
the RMNS is 
present, the higher the extent of protonization and deposition of lepton number in 
the ejecta. We leave a full investigation of the combined effects of the RMNS 
lifetime and the extent of protonization of the ejecta capture by M1 neutrino 
transport to future work.

%%\dr{It would be good to also comment on the absolute error. In the past, I
%%found that some things, like velocity distribution, are very robust, while
%%others, especially the total ejecta mass are not. This points to the fact that
%%there is some stochasticity in the mass ejection, but that the simulations are
%%``self-consistent''.}\ple{I've included some discussion on the absolute error below.}
In Tab.~\ref{tab:ejecta_properties_LR} we also show global and average ejecta 
properties for the LR simulations in our work. For reference, we also show the 
\emph{absolute} difference in the total ejecta mass predicted by each simulation,
$\delta M_{\rm ej} \equiv M_{\rm ej, tot}^{\rm SR} - M_{\rm ej, tot}^{\rm LR}$, 
while accounting for differences in the total simulation duration. Specifically, to 
calculate $\delta M_{\rm ej}$ we consider ejecta only up to the the time 
corresponding to the shortest final simulation time between LR and SR cases. 
We find that global ejecta properties 
such as the total ejecta mass differ significantly between simulations at different 
grid resolutions. Typically LR and SR simulations differ by up to approximately 
$\mathcal{O} (10^{-3}) M_\odot$, 
%%\dr{Maybe the relative difference is
%%more meaningful than the absolute difference}\ple{Fixed}
reflecting a relative difference in of 20\%-50\% in most cases. 
Depending on the model the LR simulations may either over- 
or under-estimate the amount of ejecta predicted by the SR simulations. On the other 
hand, the ejecta distributions in relevant variables 
(including $v_\infty$, $Y_{\rm e}$, and $s$) 
%%\dr{What do you mean? Maybe we should say ``as a function of Ye, vel''} \ple{Fixed} 
appear robust across grid resolutions. For instance, when comparing the 
histograms depicted in Figs.~\ref{fig:histograms_sr} and ~\ref{fig:histograms_LR}, we 
note that in all cases simulations which produce longer-lived RMNS result in 
relatively higher amounts of ejecta with $Y_{\rm e} \geq 0.4$. Moreover, the 
distributions in $v_\infty$ and $s$ appear robust across different grid resolutions. 
Significant variability in global quantities such as the total ejecta mass may 
suggest stochasticity in the mass ejection, but robustness in the ejecta 
distributions suggests self-consistency in our simulations across grid resolutions.

\bibliography{ref}

\end{document}